\documentclass{elsart}
\def\ba{\begin{eqnarray}}
\def\ea{\end{eqnarray}}
\def\ta{\tau}
\def\cal{\mathcal}
\usepackage{epsfig}
\usepackage{amsbsy}
\begin{document}
%\runauthor{Cicero, Caesar and Vergil}
\begin{frontmatter} 
\title{
Monte Carlo generator ELRADGEN 2.0 for simulation of radiative
events in elastic $ep$-scattering of polarized particles
}
\author{I. Akushevich}
\address{Duke University, Durham, NC 27708, USA}
\author{O.F. Filoti}
\address{University of New Hampshire,  Durham, NH 03824, USA}
\author{A. Ilyichev, N. Shumeiko}
\address{National
Center of Particle and High Energy Physics,
220040 Minsk, Belarus }
\begin{abstract}
The structure and algorithms of the 
Monte-Carlo generator ELRADGEN {\bf 2.0} 
designed to simulate radiative events in polarized $ep$-scattering are presented. 
The full set of analytical expressions for the QED radiative corrections is presented and discussed in detail.   
Algorithmic improvements implemented to provide faster simulation
of hard real photon events are described. 
Numerical tests show high quality of generation 
of photonic variables and radiatively corrected cross
section. 
The comparison of the elastic 
radiative tail simulated within the kinematical conditions of the 
BLAST experiment at MIT BATES shows a good agreement with experimental data.
\end{abstract}
\end{frontmatter}
\section*{PROGRAM SUMMARY}
{\it Manuscript title:}
Monte Carlo generator ELRADGEN {\bf 2.0} for simulation of radiative
events in elastic $ep$-scattering of polarized particles
\\
{\it Authors:} I. Akushevich, O.F. Filoti, A. Ilyichev, N. Shumeiko\\
{\it Program title:} ELRADGEN 2.0\\
{\it Licensing provisions:} none\\
{\it Programming language:} FORTRAN 77\\
{\it Computer(s) for which the program has been designed:} all\\
{\it Operating system(s) for which the program has been designed:} any\\
{\it RAM required to execute with typical data:} 1 MB\\
{\it Has the code been vectorised or parallelized?:} no\\
{\it Number of processors used:} 1\\
{\it Supplementary material:} none\\
{\it Keywords:} radiative corrections, Monte Carlo method,
elastic $ep$-scattering\\
{\it \underline{PACS}:} 07.05.Tp, 13.40.Ks, 13.88.+e, 25.30.Bf\\
{\it \underline{CPC Library Classification}:} \\
{\it External routines/libraries used:} none\\
{\it CPC Program Library subprograms used:}  none\\
%Catalogue identifier of previous version:* 
%Journal Reference of previous version:*
%Does the new version supersede the previous version?:*

\ \\
{\it Nature of problem:} simulation of radiative
events in polarized $ep$-scattering.\\
{\it Solution method:} Monte Carlo simulation according to the 
distributions of the real photon kinematic variables that  
are calculated by the covariant method of QED radiative correction 
estimation. 
The approach provides rather fast and accurate 
generation. 
%Reasons for the new version:*
%Summary of revisions:* 
\\{\it Restrictions:} none\\
{\it Unusual features:} none\\
{\it Additional comments:} none\\
{\it Running time:} the simulation of $10^8$ radiative events 
for $itest:=1$ takes up to 3 minutes 9 seconds on
Pentium(R) Dual-Core 2.00 GHz
processor.

\section{Introduction}

The exclusive 
photon production in lepton-nucleon scattering is the routine experimental tool in investigating the hadronic structure. 
Depending on the design of experiments, the measurements of
this process can give an access to the generalized
parton distributions \cite{Diehl,HERMES} or the generalized polarizabilities
 \cite{Gui,Dre98}.
In some cases the exclusive 
photon production appears as a background effect
to inelastic \cite{ASh,ABKR} or elastic  \cite{MaxTj} lepton nucleon scattering. 
The  last  scenario,  
{\it i.e.} the situation when the events with the real photon emission
accompany the elastic electron-proton scattering is the most advanced due to the 
infrared problem, therefore it will be in our main focus.

The set of processes contributed to the observed cross section in the next order 
of perturbation theory is referred to as the lowest order radiative corrections (RC). The basic contribution to the lowest order RC   
appears
from the square of amplitude
that only includes real photon emission from the lepton leg. 
This contribution contains the so-called large logarithm 
({\it i.e.}, the logarithm of the lepton mass) and normally is only held in the lowest order RC.  

In practice of data analysis, RC are calculated theoretically or their contribution to the observed cross sections (or asymmetries) are minimized by experimental methods.  
Due to finite detector resolution, a complete removal of  
the events with radiated hard photon(s) by pure experimental methods is not possible. 
Furthermore, the contributions of additional virtual particles and soft photon emission  
cannot be removed in principle. 
%Monte Carlo generator of radiative events is indispensable for both theoretical and experimental approaches.  
%In majority cases the lowest order RC can be calculated theoretically. 
The theoretical calculation provides with analytical expressions included the contributions of loops and 
photon emission which are infrared free after 
the procedure of the cancellation of the infrared 
divergence.
The contribution of the hard photon radiation is presented in the form of 
integrals over photon phase space. Partly the integration is performed analytically 
without additional simplifying assumptions or assumptions on specific functional 
forms describing hadronic structure. 

The pioneering approach for RC calculation in inelastic processes was suggested by 
Mo and Tsai in their seminal paper in 1969 \cite{MoTsai}. They also developed the peaking approximation allowing for analytical estimating integrals over photon angles. 
The approximation is used in many data analysis, {\it e.g. }, in ref. \cite{Ent} the electromagnetic RC in elastic $ep$-scattering was calculated 
in peaking approximation with taking into account the one-photon emission 
both from lepton and hadron legs. 

The Mo and Tsai approach requires involvement of the artificial parameter $\Delta$ separating the integration region over photon energy on parts with soft and hard photon contributions. To cancel infrared divergence analytically only leading terms in the expansion of the soft photon contribution over reciprocal of the photon energy are kept. As a result, the final expressions contain undesired dependence on the artificial parameter.  
 
Bardin and Shumeiko developed the approach \cite{BSh} for exact separation and
cancellation of infrared divergence when the final expressions for RC were completely 
free from any artificial parameters like $\Delta $. 
Using this approach, RC in polarized elastic $ep$-scattering
withing QED theory has been calculated in refs. \cite{AAM,AAIM}. Basing on these calculations the FORTRAN code
MASCARAD has been developed and successfully used for data processing of the relevant parity conservation experiments
\cite{JLep1,JLep2}. 
Other approaches were also used for RC calculation in elastic $ep$-scattering. Thus, the total lowest order RC 
(both to lepton and hadron legs) was also calculated in 
\cite{MaxTj} with soft photon approximation and the method of electron structure functions
suggested in the work \cite{ESF} was also applied for estimation of RC to elastic $ep$ scattering
\cite{ESF1,ESF2}.

The use of realistic detector geometry requires essentially complicated 
integration over the real photon phase space. As a result, the researchers come to 
the necessity of using the Monte Carlo technique which constitutes a complementary approach to the theoretical calculations of RC using respective codes such as MASCARAD. 
The Monte Carlo generators 
for simulation of radiative events have been developed for many specific processes and intensively used in data analysis. Thus, the Monte 
Carlo generator RADGEN \cite{RADGEN} for simulation of radiative
events in inclusive deep inelastic scattering has been developed on the basis of the FORTRAN code POLRAD  \cite{POLRAD}. The Monte Carlo generator  MERADGEN \cite{meradgen} 
for  simulation of radiative
events in  M\o ller scattering  appeared on the  base of FORTRAN code MERA \cite{mera}.

In this paper we present and describe in detail the latest version {\bf 2.0} of the Monte Carlo generator 
ELRADGEN. The prototype of the code~\cite{elradgen} dealt with the simulation
of real hard photon emission as a background effect in the unpolarized 
elastic electron-proton scattering. 
The present version {\bf 2.0} is extended  on 
the initial polarized particles: longitudinally polarized electron
and arbitrary polarized proton. The theoretical background for the developments is presented in ref. \cite{AAM}.

The paper is organized as follows.
Section \ref{kin} describes the kinematics
of the investigated process and the generation method.
The different contributions to the 
lowest order RC and the multi-soft photon emission are presented and discussed in Section \ref{xxsect}. The brief structure of the code and the input-output datafiles are described 
in Section \ref{sstruct}. 
Test runs, comparison with MASCARAD, and numerical comparison of the simulated and measured cross sections of the radiative tail from elastic peak in the 
BLAST experiment are presented in  Section \ref{testsdata}.
Conclusions and final remarks are given 
in Section \ref{cconc}. The four-momenta reconstruction formulae, explicit expressions
for the lepton and target polarization vectors, some lengthy formulae
for RC, and test outputs are given in Appendices.

\begin{figure}[t]
\vspace*{60mm}
\hspace*{-8mm}
\unitlength 1mm
\begin{tabular}{ccccc}
\begin{picture}(20,20)
\put(-13,0){
\epsfxsize=65mm
\epsfysize=7cm
\epsfbox{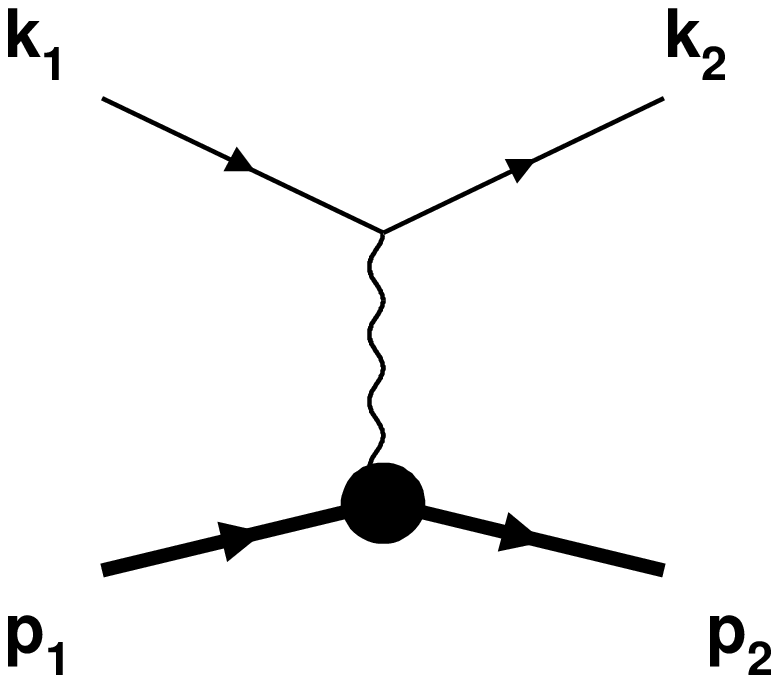}
\put(-35,48){\mbox{(a)}}
}
\end{picture}
&
\begin{picture}(20,20)
\put(-10,0){
\epsfxsize=65mm
\epsfysize=7cm
\epsfbox{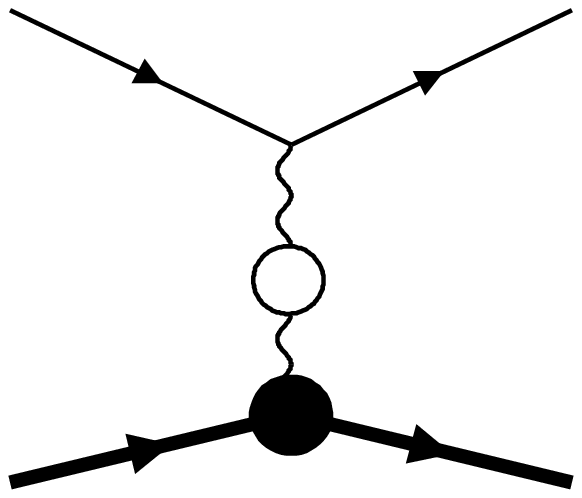}
\put(-35,48){\mbox{(b)}}
}
\end{picture}
&
\begin{picture}(20,20)
\put(-7,0){
\epsfxsize=65mm
\epsfysize=7cm
\epsfbox{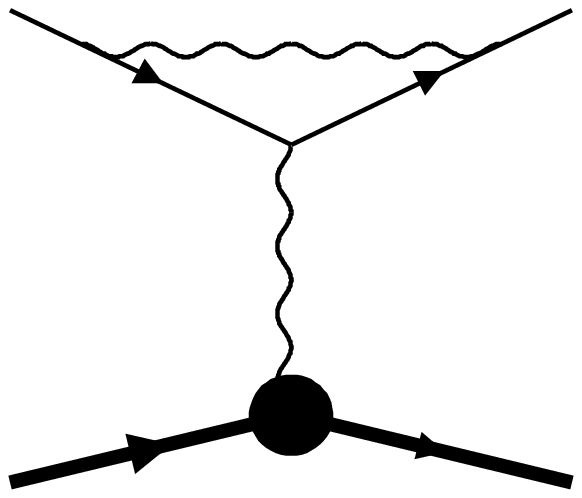}	
\put(-35,48){\mbox{(c)}}
}
\end{picture}
&
\begin{picture}(20,20)
\put(-4,0){
\epsfxsize=65mm
\epsfysize=7cm
\epsfbox{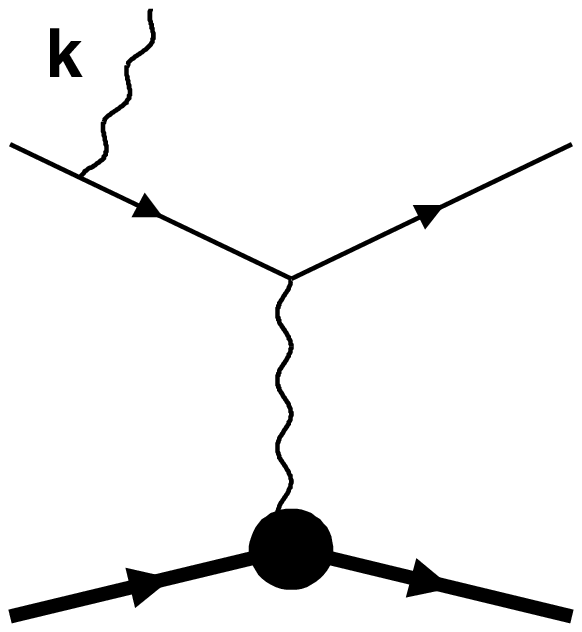}
\put(-35,48){\mbox{(d)}}
}
\end{picture}
&
\begin{picture}(20,20)
\put(-1,0){
\epsfxsize=65mm
\epsfysize=7cm
\epsfbox{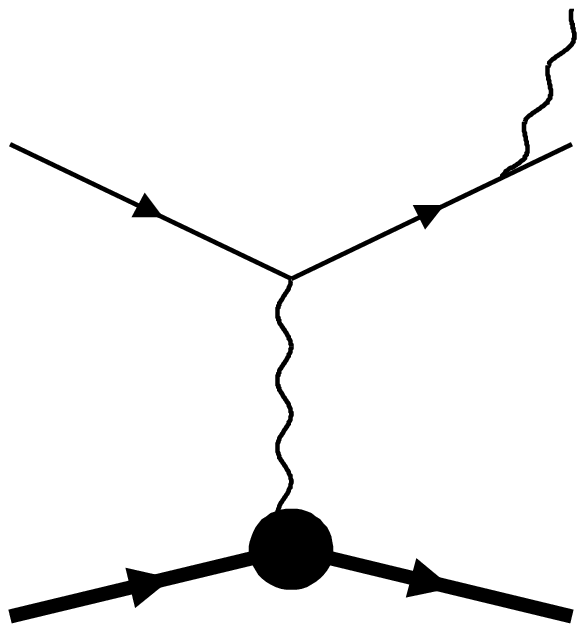}
\put(-35,48){\mbox{(e)}}
}
\end{picture}
\end{tabular}
\vspace{-45mm}
\caption{\label{fg}
\protect\it
Feynman graphs contributing to radiatively corrected cross sections
of elastic lepton-nucleus scattering:
Born (a), additional virtual particles (b,c) and real photon emission (d,e)
contributions.}
\vspace{5mm}
\end{figure}

\section{Kinematics and Method of Generation}
\label{kin}
The lowest order (Born) (Fig.\ref{fg}~(a)) as well as the additional  
virtual particle (Fig.\ref{fg}~(b,c))
contributions to the polarized
elastic lepton-nucleon scattering
\ba
e(k_1,\; \xi _L)+p(p_1,\; \eta )\longrightarrow e'(k_2)+p'(p_2)
\label{main}
\ea
($k_1^2=k_2^2=m^2$, $p_1^2=p_2^2=M^2$)
can be described by the following three variables:
\ba
Q^2=-q^2=-(k_1-k_2)^2,\qquad S=2k_1p_1,\quad  \phi,
\label{k0}
\ea
where $\phi$ is the azimuthal angle between the scattering
plane ${\bf (k_1,k_2)}$ and the ground level. The Lab system is used 
with OZ axis along the beam direction and plane OZX
parallel to the ground  level. The explicit expressions of polarization vectors ($\xi _L$ and $\eta $) and four-momenta reconstructed in the lab system are presented in
\ref{fv}.

 The description of the phase space
of the radiative process (Fig.\ref{fg}(d,e))
\begin{equation}
e(k_1,\; \xi _L)+p(p_1,\; \eta )\longrightarrow e'(k_2)+p'(p_2)+\gamma (k),
\end{equation}
($k^2=0$) requires
three new kinematic variables: a virtual proton transfer momentum squared
$t=-(k_1-k_2-k)^2$,
the inelasticity $v=(p_2+k)^2-M^2$, and the azimuthal angle $\phi_k$
between the planes ${\bf(q,k)}$ and  ${\bf
(k_1,k_2)}$.
This set of variables defines the four-momenta of all final
particles.

The simulation of radiative events requires an additional definition of the lowest bound of
the photon energy (or another respective quantity, inelasticity $v_{min}$ in our case) 
separating the photon phase space into the region of soft and hard photons. 
Only hard photons need to be simulated while soft photons cannot be simulated because of the infrared 
divergence. The observed cross section can be presented in terms of two positively definite parts:
\begin{equation}
\sigma _{obs} =
\sigma _{rad}(v_{min})
+
\sigma _{BSV}(v_{min}). 
\label{sobr}
\end{equation}
The first term, $\sigma _{rad}(v_{min})$, describes the cross section with an additional hard photon emitted, 
and the second, $\sigma _{BSV}(v_{min})$, 
contains the contributions of the Born cross section, soft-photon emission, and virtual corrections.  
Here and later we define $\sigma \equiv d\sigma /dQ^2 d\phi$. Note that $\sigma _{obs}$ does not depend on 
$v_{min}$ while terms $\sigma _{BSV}$ and $\sigma _{rad}$
do. 

The strategy for simulation of one event can be defined 
in a standard way \cite{elradgen}:
\begin{itemize}
\item For the fixed initial energy, $Q^2$, the angle $\phi$, and
the missing mass square resolution $v_{min}$,  
the two positively-definite contributions to the observed (radiative-corrected) cross section
$\sigma _{obs}$,
$\sigma _{rad}(v_{min})$,
and
$\sigma _{BSV}(v_{min})$
are calculated separately.

\item The corresponding channel of scattering  ({\it i.e.}, BSV or radiative process) 
is simulated for this event in accordance with partial 
contributions of these two positive parts
into the total cross section. More specifically, the channel of scattering is 
simulated in accordance with the Bernoulli 
trial where the probability of ``success'' 
({\it i.e.}, radiative channel) is calculated as a ratio of the radiative part of the cross section
 to the total cross section.  
\item For the radiative event the kinematic variables $t$, $v$ and
$\phi_k$
are simulated in accordance with their calculated distributions. 
The distributions of $v$ and $\phi _k$ 
are conditional ({\it e.g. }, $v$ is simulated conditionally on $t$, and $\phi_k$ is simulated conditionally 
on $t$ and $v$). The explicit expressions for the probability densities of these variables are defined
by eqs. \ref{roo}.
\item The four--momenta of all final particles in a required reference frame
are calculated.
\end{itemize}

The initial values of $Q^2$ (and $\phi$) can be non-fixed but externally simulated 
according to a probability distribution (for example,
the Born cross section). If the $Q^2$ distribution is simulated over the Born cross section, then the realistic
observed
$Q^2$ distribution is calculated as sum of weights computed as ratios of the total and Born cross sections for each simulated event. If the observed cross section is used for the simulation of $Q^2$, then reweighting is not required.

\section{Explicit expressions for 
$\sigma _{rad}(v_{min})$ and
$\sigma _{BSV}(v_{min})$ 
}
\label{xxsect}

The 
analytical expressions for the lowest order RC on which the ELRADGEN is based, were obtained in ref. 
\cite{AAM} (see eqs. (50) and (51)). The result for the observed cross section can be formally outlined
as $\displaystyle \sigma_{obs}=(1+\delta )\sigma_0+C\int \frac{dv}v [\sigma_R(v)-\sigma_0]$, 
where $C$ is a kinematic coefficient proportional to $\alpha$ and the quantity 
$\sigma_R(v )$ is proportional to the bremsstrahlung cross section ($\sigma_R(0)=\sigma_0$). 
This expression does not reproduce the form of eq. (\ref{sobr}), because the term with the integral 
is not positively definite and the term with $\sigma_R(v )$ cannot be separated because 
it is singular for $v \rightarrow 0$. Instead, the following transformation of this term was used:
\begin{eqnarray}
&&\int {dv \over  v} (\sigma_R( v)-\sigma_0)=
\int {dv \over v} \sigma_R(v)\theta(v-v_{min})
\nonumber
\\&&\qquad\qquad
-\int {dv \over v} \sigma_0 \theta(v-v_{min})
+\int {dv \over v} [\sigma_R(v)-\sigma_0] \theta(v_{min}-v).
\label{illustr}
\end{eqnarray}
The first term in (\ref{illustr}) represents the contribution of hard photons, {\it i.e.}, 
with inelasticity above $v_{min}$. This term is positively definite and it is used as 
$\sigma _{rad}(v_{min})$ in (\ref{sobr}). Its structure and explicit expressions are discussed 
in Section {\ref{brems}}. The second term admits the analytic integration resulting in correction 
$\delta^{add}(v_{min})$. This term (as well as the third term in the eq.~(\ref{illustr}) discussed in 
Section \ref{sigmaadd}) contributes to the $\sigma _{BSV}(v_{min})$ that  represents the part of the observed 
cross section not contained in the contributions of radiated photons with inelasticity above       
$v_{min}$.  

\subsection{$BSV$ cross section}

The $BSV$-part of observed cross section includes the Born cross section (Fig.\ref{fg}~(a)), loop effects
(Fig.\ref{fg}~(b,c)) and the contribution of soft photons.
The latter is restricted by the inelasticity value $v<v_{min}$:
\ba
\sigma _{BSV}(v_{min}) &=&
(1+ \delta_{VR}+\delta_{vac}^l+\delta_{vac}^h)
e^{\delta_{inf}}\sigma _{0}
+\delta^{add}(v_{min})
\sigma _{0}
\nonumber \\&&
+\sigma^{add}_{R}(v_{min}).
\label{BSV}
\ea
The Born contribution to the cross section reads:
\begin{equation}
\sigma _0=\frac{\alpha ^2}
{S^2 Q^4}\sum_{i=1}^4 \theta_i ^B{\mathcal  F}_{i}(Q^2).
\end{equation}
 The kinematic coefficients
$\theta _B$ are presented in \ref{exp}. The structure functions ${\mathcal  F}_i$ are the squared 
combinations of the electric
and
magnetic elastic form factors:
\ba
&&{\mathcal  F}_1(Q^2)=4\ta _pM^2G_M^2(Q^2),\;\;
{\mathcal  F}_2(Q^2)=
4M^2\frac {G_E^2(Q^2)+\ta _p G_M^2(Q^2)}{1+\ta _p},
\nonumber \\
&&{\mathcal  F}_3(Q^2)=-2M^2G_E(Q^2)G_M(Q^2),\;\;
\nonumber \\[2mm]
&&{\mathcal  F}_4(Q^2)=
-M^2G_M(Q^2)\frac {G_E(Q^2)-G_M(Q^2)}{1+\ta _p}
\ea
with $\ta _p=Q^2/4M^2$. 

The factorizing corrections in the first term of (\ref{BSV}) describe the effects 
of loops and soft-photon emission. The correction $\delta_{inf}$ comes from the emission 
of soft photons, the $\delta_{VR}$ appears as a result of an infrared cancellation of real 
(Fig.\ref{fg} (d,e)) and virtual (Fig.\ref{fg} (c)) photon contribution. 
The explicit expressions for them are:
\ba
\delta_{inf}&=&{\alpha\over \pi}\Bigl( \log \frac{Q^2}{m^2}-1 \Bigr) \log \frac{v_{max}^2}{S(S-Q^2)},
\nonumber \\
\delta_{VR}&=&{\alpha\over\pi}\biggl(\frac{3}{2}\log \frac{Q^2}{m^2}	-2-\frac{1}{2}\log^2\frac{S}{S-Q^2}
+{\rm Li}_2\biggl( 1-\frac{M^2Q^2}{S(S-Q^2)}\biggr )
\nonumber \\&&
-\frac{\pi^2}{6}\biggr),
\ea
where ${\rm Li}_2$ is the Spence function.

The effect of vacuum polarization by leptons (hadrons) depicted on Fig.\ref{fg}~(b)
is described by  $\delta^l _{vac}$ ($\delta^h _{vac}$).  
The explicit expression for $\delta^l _{vac}$ is defined by eq.~(21) of ref. \cite{ASh}
while the fit for $\delta^h _{vac}$ has been taken from \cite{hadvac}. 

The term $\delta^{add}(v_{min})\sigma _{0}$ in the R.H.S. of (\ref{BSV}) contains the correction coming from the second term in (\ref{illustr}). 
\begin{eqnarray}
\delta ^{add}(v_{min})&=&
-\frac{2\alpha }{\pi}
\Bigl(\log \frac {Q^2}{m^2}-1\Bigr)
\log \frac {v_{max}} {v_{min}} .
\end{eqnarray}
The last term in (\ref{BSV}) is discussed in Section \ref{sigmaadd}.

\subsection{Bremsstrahlung cross section}
\label{brems}
\begin{figure}[t]
\unitlength 1mm
%\vspace*{10mm}
\begin{picture}(160,80)
\put(20,0){
\epsfxsize=10cm
\epsfysize=10cm
\epsfbox{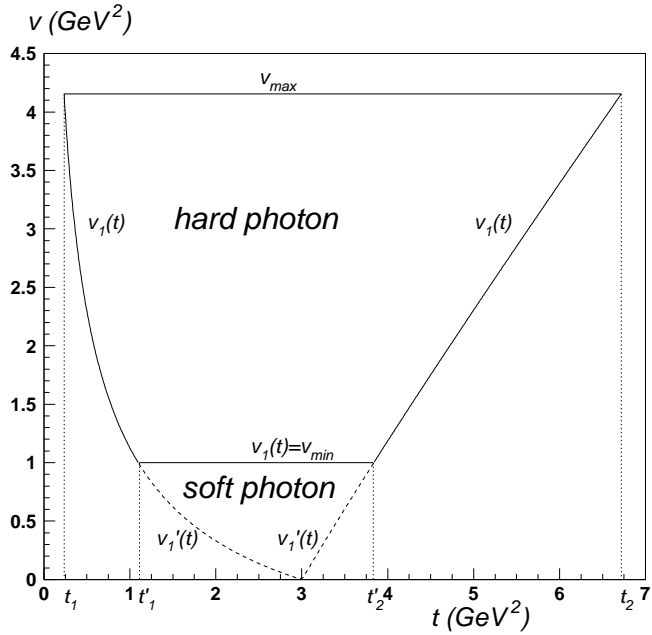}
}
\end{picture}
\label{fig1}
\vspace*{-10mm}
\caption{\label{kinreg}
The region of integration over $v$- and $t$-variables
for JLab kinematics ($Q^2=3$ GeV$^2$, $S=7.5$ GeV$^2$).
The line $v=v_{min}$ splits it into hard (solid lines) and soft
(dashed lines) real photon regions.}
\end{figure}

Since the structure 
functions  
depend only on $t$, and therefore integrals over other variables 
({\it i.e.}, $v$ and $\phi _k$) can be evaluated analytically or numerically with high precision, 
a reasonable sequence of integration variables is chosen such that integration over $t$ is external. 
This approach allows us to speed up the generation of radiative events. 
The radiative photon phase space for $t$- and $v$-variables are
presented on Fig.~\ref{kinreg}. 
It is separated into hard and soft photon
emission by the line $v=v_{min}$. The cross section of hard-photon bremsstrahlung is
\begin{equation}
\sigma ^{rad}(v_{min})
= -{\alpha^3 \over 4\pi S^2}
\int\limits_{t_1}^{t_2}dt
\sum_{i=1}^4
{{\mathcal  F}_i(t) \over t^2}
\theta_{i}^R(v_1,v_{max}).
\label{tt}
\end{equation}
The quantities $\theta_{i}^R(v_1,v_{max})$ result from the integration over inelasticity $v$. 
Their arguments 
correspond to the limits of integration:
\ba
\theta _i^R (v_1,v_{max})
&=&
\sum_{j=1}^{k_i}
\int \limits _{v_1}^{v_{max}}dv
R^{j-3}\theta_{ij}^R (v ).
\label{thr000}
\ea
Here $R=Q^2+v-t$, and the upper sum limits are defined as $k_i=(3,3,4,5)$. Accordingly, the quantities $\theta_{ij}^R (v )$ result from the integration over $\phi_k$:
\ba
\theta_{ij}^R (v )
&=&
\int \limits _0^{2\pi}d\phi _k \
\theta_{ij}^R (v, \phi _k ).
\label{thr0}
\ea
The set of 
quantities $\theta^R $ is defined in \ref{exp}.  

Kinematical bounds are defined as
\begin{eqnarray}
v_1&=&\max\{ 
\frac{(t-Q^2)(\sqrt{t}-\sqrt{4M^2+t})}{2\sqrt{t}}, 
\frac{(t-Q^2)(\sqrt{t}+\sqrt{4M^2+t})}{2\sqrt{t}}, 
\nonumber \\&&\qquad
v_{min}\},\;
\nonumber \\[2mm]
v_{max}&=&\frac {2Q^2(S^2-4M^2m^2-Q^2(S+m^2+M^2))}{Q^2(S+2m^2)
+\sqrt{Q^2(S^2-4M^2m^2)(Q^2+4m^2)}}
\nonumber \\
&\approx& 
S-Q^2-\frac{M^2Q^2}S,
\nonumber \\
t_{1,2}
&=&\frac{2M^2Q^2+v_{max}\left (Q^2
+v_{max}\mp
\sqrt{(Q^2+v_{max})^2+4M^2Q^2}\right )}{2(M^2+v_{max})}.
\end{eqnarray}

The 
probability distributions used for simulation of 
the photonic variables are obtained using (\ref{tt}) and (\ref{thr0}):
\ba
\rho(t)&=&\frac 1{N_t}\sum_{i=1}^4\frac{{\cal F}_i(t)}{t^2}\theta _i^R(v_1,v_{max}),
 \;\;\;\;
N_t=\sum_{i=1}^4\int\limits_{t_1}^{t_2}dt\frac{{\cal F}_i(t)}{t^2}\theta _i^R(v_1,v_{max}),
\nonumber\\
\rho(v| t)&=&\frac 1{N_v}
\sum_{i=1}^4\sum_{j=1}^{k_j}{\cal F}_i(t)\theta _{ij}^R(v)R^{j-3},
 \;\;
N_v=\sum_{i=1}^4{\cal F}_i(t)\theta _i^R(v_1,v_{max}),
\qquad
\nonumber\\
\rho(\phi _k| v,t)&=&\sum_{i=1}^4\sum_{j=1}^{k_j}\frac {\displaystyle
{\cal F}_i(t)\theta _{ij}^R(v,\phi_k)R^{j-3}}{N_{\phi _k} }
,
\;
N_{\phi _k}=\sum_{i=1}^4\sum_{j=1}^{k_j}{\cal F}_i(t)\theta _i^R(v)R^{j-3}.
\;\;
\label{roo}
\ea

\subsection{Contribution of $\sigma _R ^{add}(v_{min})$}
\label{sigmaadd}
The contribution of $\sigma _R^{add}(v_{min})$ can be presented as an
integral over the soft-photon region in Fig.~\ref{kinreg}:
\ba
\sigma _R^{add}{(v_{min})}
&=& -{\alpha^3 \over 4\pi S^2}\int\limits_{t'_1}^{t'_2}
dt \int\limits_{v'_1}^{v_{min}}
dv \sum_{i=1}^4 \Biggl [
\sum_{j=2}^{k_j}R^{j-3}\theta_{ij}^R (v)
{{\mathcal  F}_i(t) \over t^2}
\nonumber \\&&+
\frac 1{R^2}
\Biggl (\theta_{i1}^R (v)
{{\mathcal  F}_i(t) \over t^2}
-
4 \theta^B_{i}
F_{IR}(v){{\mathcal  F}_i(Q^2) \over Q^4}\Biggr)
\Biggr ].
\label{t}
\ea
The limits of integration over variables $t$ and $v$ read:
\ba
v_1'&=&\max \{
\frac{(t-Q^2)(\sqrt{t}-\sqrt{4M^2+t})}{2\sqrt{t}},\; 
\frac{(t-Q^2)(\sqrt{t}+\sqrt{4M^2+t})}{2\sqrt{t}} 
\},\;
\nonumber \\
t_{1,2}'
&=&\frac{2M^2Q^2+v_{min}\left (Q^2
+v_{min}\mp
\sqrt{(Q^2+v_{min})^2+4M^2Q^2}\right )}{2(M^2+v_{min})}.
\ea

The infrared divergences could occur in the limit $v_1'\to 0$  ({\it i.e.} at $t\to Q^2$)
in the terms containing $R^{-2}$.
However,
one can see that $\sigma _R^{add}{(v_{min})}$ is infrared-free. 
Indeed, taking into account Eqs.~(\ref{thr0}), (\ref{thb}), (\ref{th13}), and (\ref{th41}),
in the limit $v_1'\to 0$ we have
\ba
\lim_{v\to 0}\theta_{i1}^R (v)
{{\mathcal  F}_i(t) \over t^2} =
4 \theta^B_{i}
F_{IR}(0){{\mathcal  F}_i(Q^2) \over Q^4}.
\label{thv0}
\ea
This cancels one degree of $R$. The second degree of $R$ cancels because the integration region is collapsed into a point within this limit. 

\section{The structure of the program and input-output data}
\label{sstruct}
\begin{figure}[t]
\unitlength 1mm
%\hspace*{-2cm}
%\vspace*{5cm}
\begin{picture}(80,80)
\put(-5,-40){
\epsfxsize=15cm
\epsfysize=19cm
\epsfbox{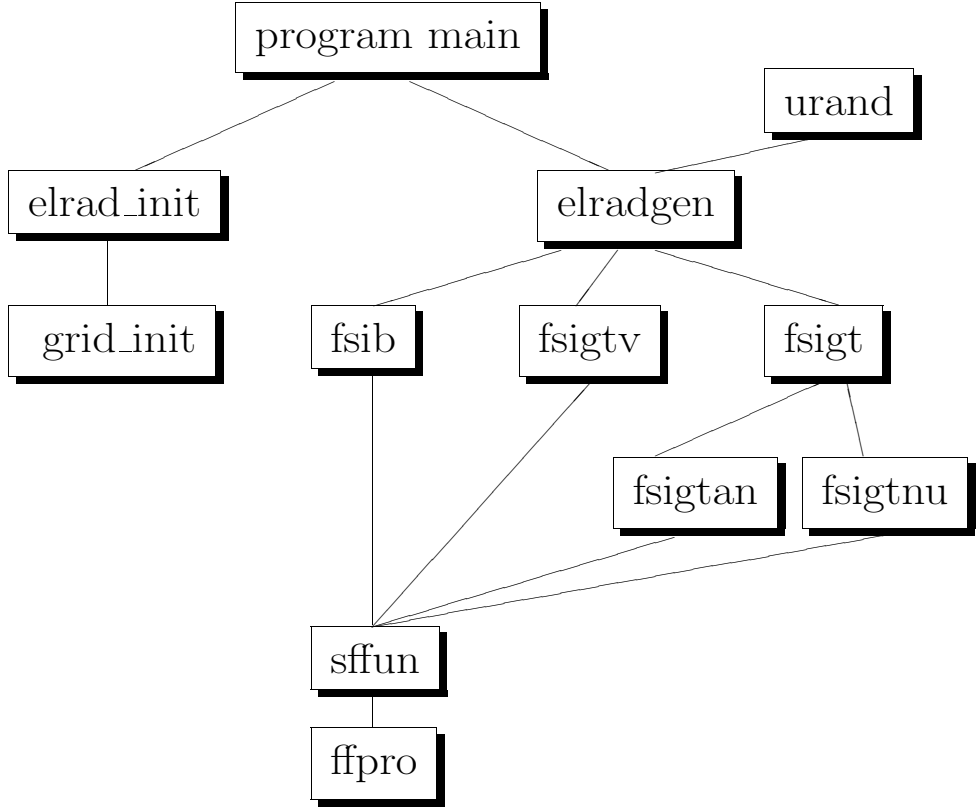}
}
\end{picture}
\vspace*{33mm}
\caption{\label{str}
The structure of the program ELRADGEN 2.0
}
\end{figure}

\subsection{The structure of the program}
The set of files included in the package ELRADGEN 2.0 contains 
 three FORTRAN files 
({\bf elradgen.f}, {\bf run.f}, {\bf test.f}), six INCLUDE files
({\bf const.inc}, {\bf grid.inc}, {\bf output.inc}, {\bf par.inc}, {\bf pol.inc},
{\bf test.inc}), two data files ({\bf rnd.dat}, {\bf test.dat}), and one {\bf Makefile}. 
No installation is required for this code.  

The {\bf elradgen.f} is a source code of 
the Monte Carlo generator ELRADGEN 2.0. It contains the set of functions and subroutines for simulation of a single event. The loop over simulated events as well as initialization of constants requires coding in the external program. Two versions of such external programs are given in files {\bf run.f} and {\bf test.f}. The file {\bf run.f} is a typical external program for simulation of an event with fixed $Q^2$ and $\phi$. The file {\bf test.f} is designed to run several tests discussed below. 

The structure of the code is illustrated in Fig.~\ref{str}: 
\begin{itemize}
\item
{ \bf program main } is a sample of an external program that invokes ELRADGEN (in our case it is the program included in {\bf run.f} and {\bf test.f});
\item
{ \bf elrad$\_$init } defines all constants (such as beam energy and polarization degrees)
which are necessary for generation;
\item
{ \bf grid$\_$init } prepares the grids for generation of photonic
kinematic variables; 
\item
{ \bf elradgen } is a main subroutine governing the simulation of an event;
\item
{ \bf urand } is a generator of uniformly distributed random numbers;
\item
{ \bf fsib } calculates the Born cross section; 
\item
{\bf fsigt } invokes one of the subroutines {\bf fsigtan } 
or {\bf fsigtnu} to calculate the cross section $d\sigma /dt$;
\item
{\bf fsigtan } calculates the analytical cross section $d\sigma /dt$ for unpolarized scattering;
\item
{\bf fsigtnu } calculates the cross section $d\sigma /dt$ with numerical integration
over variable $v$ for polarized scattering;
\item
{\bf fsigtv } calculates the analytical cross sections
$d\sigma /(dtdv)$ and $d\sigma /(dtdvd\phi _k)$;
\item
{\bf ffpro } is a model for elastic form factors.
 
\end{itemize}

The six INCLUDE files are:
\begin{itemize}
\item
{\bf const.inc} includes all necessary constants, {\it e.g. }, the fine electromagnetic constant, the proton and lepton masses;
\item
{\bf grid.inc} includes   nets of bins for simulation of the three photonic variables;
\item
{\bf output.inc} contains variables governing the form of output as discussed below
\item
{\bf pol.inc} includes quantities which describe the polarization state (defined in \ref{pv});
\item
{\bf test.inc} includes variables and nets of bins required for test run;
\item
{\bf par.inc} includes variables required for calculation of $\sigma ^{add} _R (v_{min} )$.
 
\end{itemize}

The file {\bf rnd.dat} includes an initial integer  for the flat generator {\bf urand}, and  
{\bf test.dat} is an example of output data for a test run (when {\bf test.f} is used as an external program); the results of  different test are presented in  
 \ref{tout}.

The commands ``make'' or ``make test'' need to be run for creating the executable file for the simulation or for the test runs, respectively.   

\subsection{Input-output data}
Input data in ELRADGEN 2.0 are set up in {\bf program main} 
of {\bf run.f} or {\bf test.f}. Majority of them are transferred to the main program through parameters in the {\bf subroutine elradgen}. They are:
\begin{itemize} 	
\item
{\bf ebeam} is an energy of electron beam;
\item
{\bf q2} is a virtual photon momentum squared $Q^2$;
\item
{\bf phi } is an azimuthal angle between the scattering plane and the ground level; 
\item
{\bf vvmin} is a missing mass square resolution $v_{min} $ for separation of radiatively corrected
cross section into radiative and BSV parts;
\item
{\bf vcut} is a cut-off quantity $v_{cut}$ that allows to exclude the simulation of hard photons above $v_{cut}$.
\end{itemize} 
The last variable provides the opportunity to exclude simulation of events with inelasticity above a predetermined level. This could be convenient when simulation is performed for experimental design, when hard real photon are
removed from experimental data by putting a cut on the missing mass of the undetectable particle.  

The quantities describing the polarization characteristics of beam and target (defined in \ref{pv}) are transferred to the code through the common block {\bf pol} containing four variables: 
 i,ii) $plrun$ and $pnrun$, the polarization degrees of the lepton beam $P_L$ and target $P_N$, iii) $thetapn$,  the angle $\theta _{\eta }$ between 3-vectors of the target polarization 
$\boldsymbol  \eta $ and initial lepton momentum $\bf k_{1}$, and iv) $phipn$, the angle $\phi _{\eta }$ between
OZX and ($\bf k_{1}$, $\boldsymbol \eta$) planes. 

One additional variable $itest$ governs the form of the output. If $itest \neq 0$, all output information is printed to the file {\bf test.dat}. 
If $itest=0$, the output data are collected in two common blocks 
of the file {\bf output.inc}: 

	common/variables/tgen,vgen,phigen,weight,ich

and
	
	common/vectors/vprad,phrad
	
Here $tgen$, $vgen$, and $phigen$ are the generated photonic 
variables $t$, $v$, and $\phi _k$, respectively,
$weight$ is a ratio of the observable cross section to 
the Born one, variable $ich$ shows whether 
the scattering channel is radiative ($ich=1$) 
or BSV ($ich=0$).  
The quantities $vprad=p_2-p_1$ and $phrad:=k$ are four-momenta of virtual and real photons defined in the Lab system.

For BSV events, $vgen=0$, $tgen=Q^2$,
$\phi _k=0$,  $phrad=0$ and $vprad=k_1-k_2$.

\section{Numerical tests and comparison with experimental data}
\label{testsdata}
Below we describe three types of numerical experiments allowing: i) to crosscheck some key distributions and parameter estimates in ELRADGEN, ii) to investigate issues related to a possible dependence of simulated cross sections on $v_{min}$, and iii) to perform a comparison with data collected in the BLAST experiment.   

\subsection{Tests implemented in ELRADGEN}  
\label{ttest}
There are five tests implemented in the program. The first three deal with checking to which extent the simulated distributions on 
photonic variables $t$, $v$, and $\phi _k$ correspond to analytical 
probability distributions given in eq. (\ref{roo}).

\begin{figure}[!t]
\unitlength 1mm
%\hspace*{-2cm}
%\vspace*{-10mm}
\begin{tabular}{cc}
\begin{picture}(80,80)
\put(-5,0){
\epsfxsize=7.5cm
\epsfysize=7.5cm
\epsfbox{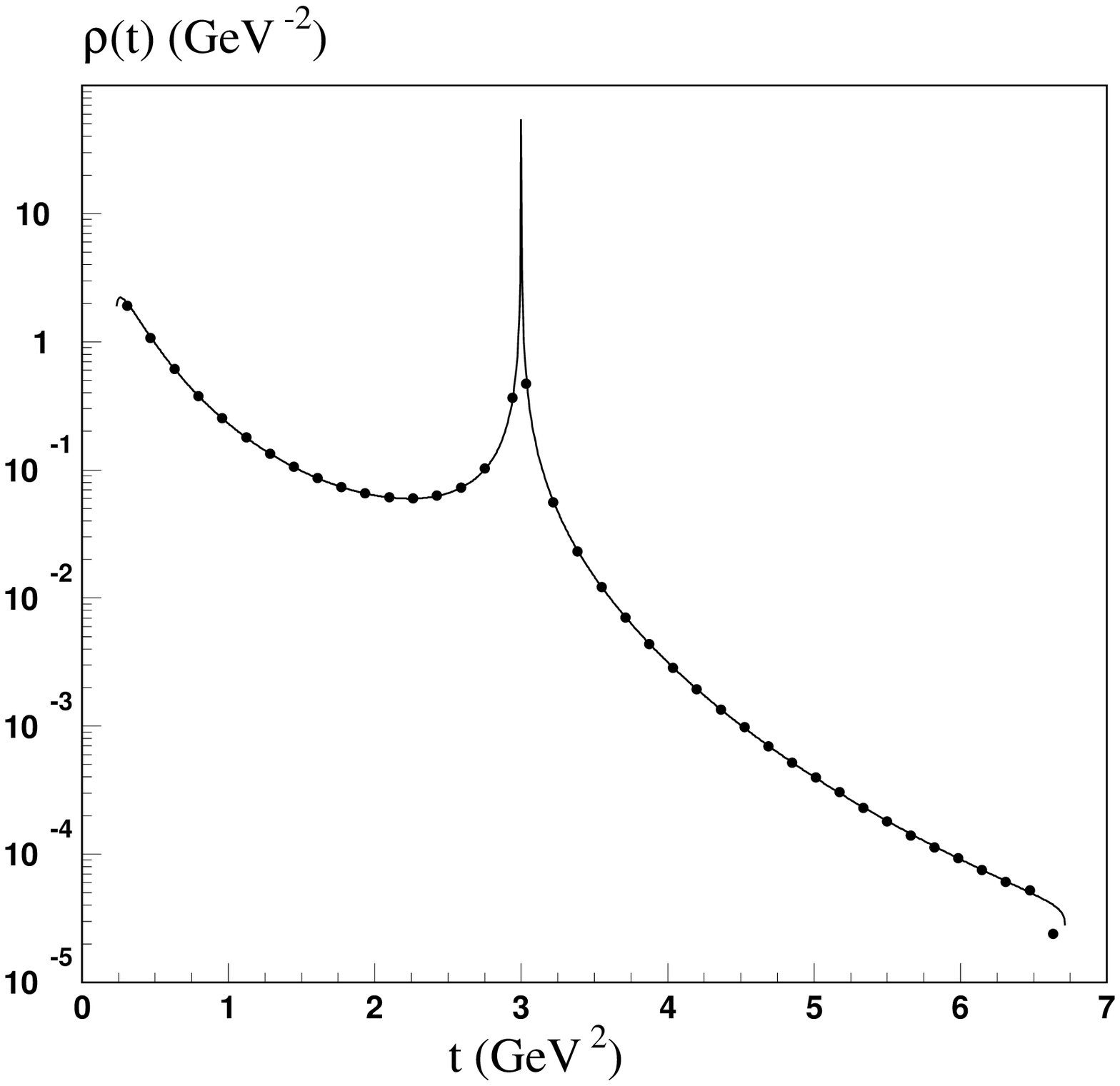}
}
\end{picture}
&
\begin{picture}(80,80)
\put(-15,0){
\epsfxsize=7.5cm
\epsfysize=7.5cm
\epsfbox{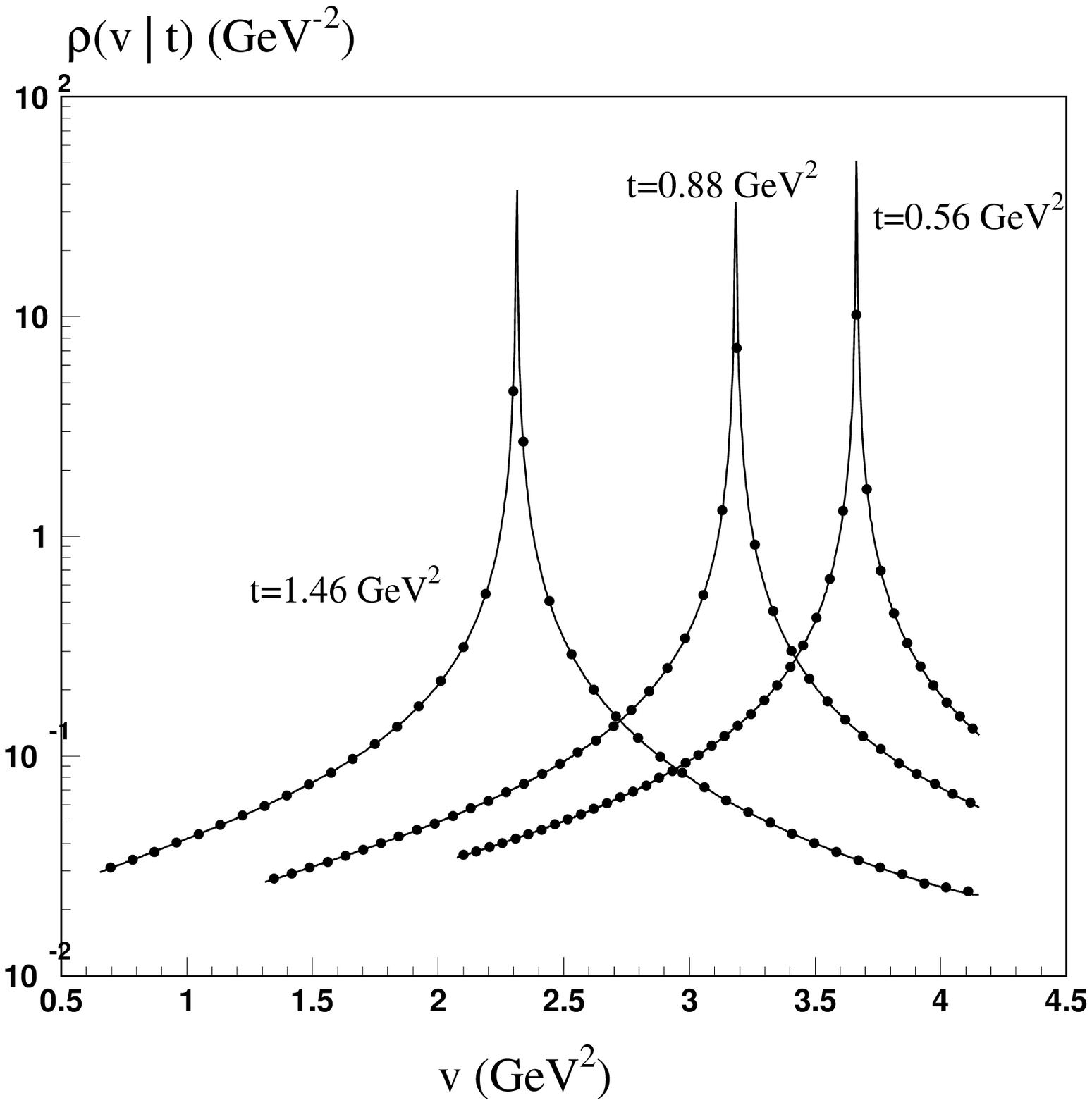}
}
\end{picture}
\\[-15mm]
\begin{picture}(80,80)
\put(-5,0){
\epsfxsize=7.5cm
\epsfysize=7.5cm
\epsfbox{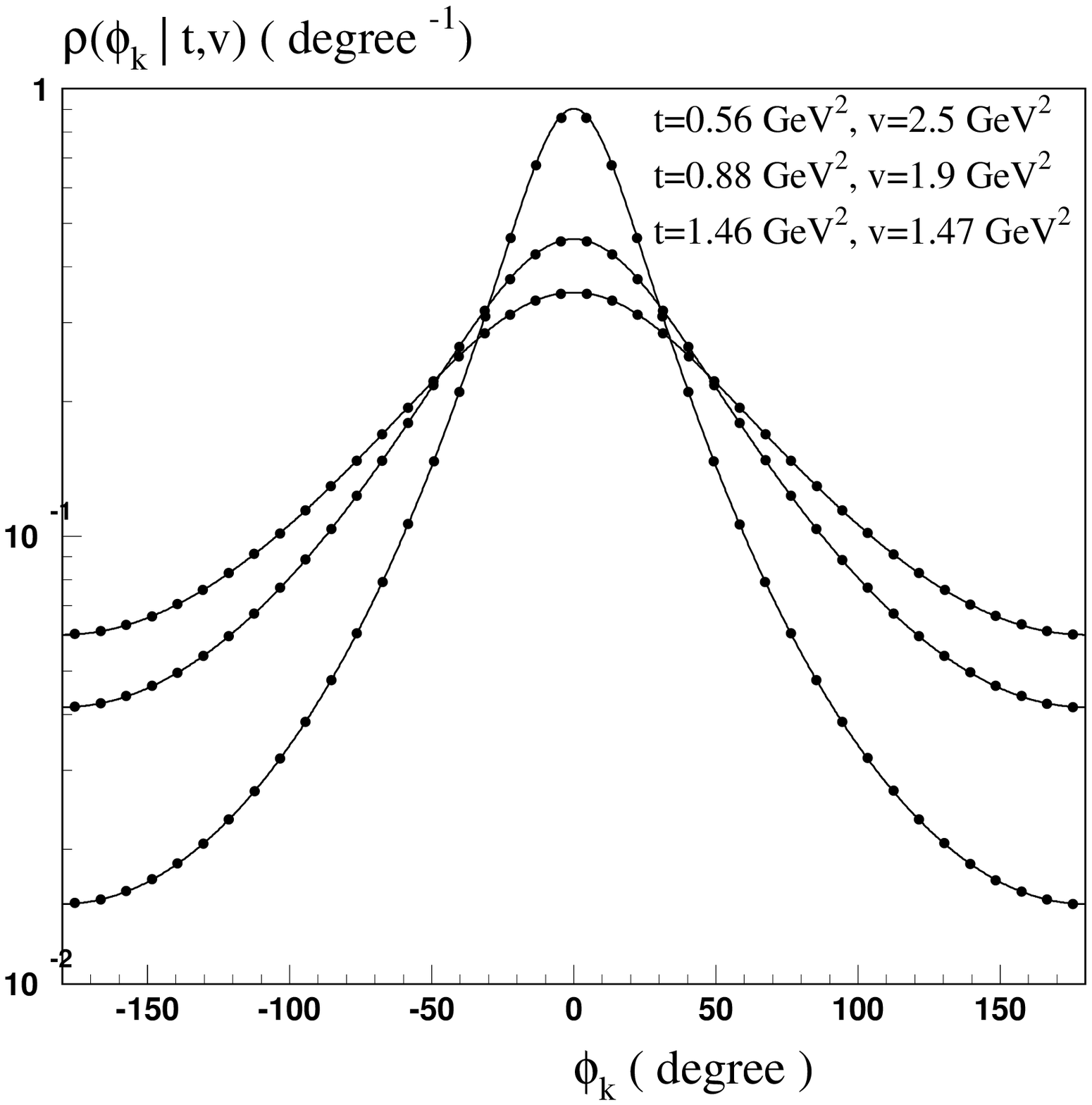}
}
\end{picture}
\end{tabular}
\caption{Histogram (points)
and corresponding probability densities  (solid lines)
for variables describing the
 exclusive real hard photon production in polarized
electron proton scattering
at JLab kinematic conditions 
($E_{beam}=4$~GeV, $Q^2=3$ GeV$^2$, 
) for transverse polarized proton
($\theta_{\eta}=90^0$) with 
$\phi=\phi_{\eta}$, $P_LP_N=-1$
and $v_{min}=10^{-2}$~GeV$^2$. 
}
\label{dist}
\end{figure}

Fig. \ref{dist} presents the $t$-, $v$-, and 
$\phi _k$-distributions calculated numerically and generated by ELRADGEN
under JLab kinematic conditions for a transverse polarized target.
The theoretical and simulated distributions of this case (as well as for unpolarized and longitudinally polarized targets) are almost identical.

The sharp peaks in the $t$-distribution coming from the collinear 
singularities, {\it i.e.}, from the kinematical region where the real photon is emitted along either
the initial or the final
lepton. After integration over the inelasticity $v$, these two singularities are situated
near $t=Q^2$ and are only slightly different.  

The peaks on the plots of the $v$-distributions correspond to the collinear 
singularities as well. 
Since the variable $t$ is external, the $v$-distribution is conditional on $t$, and therefore only one 
peak corresponding to either the initial or the final electron appears for each $v$-distribution. 
%Mathematically the peak appears when statement means that the one 
%of the leptonic propagator in Fig. \ref{fg} (d, e) reaches its maximum value
%which corresponds to the situation when the quantities $C_{1,2}$ in eq. %(\ref{bc})
%are proportional to $m^2$ \footnote{poka ne znaju nado li nam ehto %predlozhenie}.  

Finally, the $\phi _k$-distributions show that most of the photons are emitted
in the scattering plane. 

For generation of the three types of distributions, one has to set
$itest=1$, $itest=2$, or $itest=3$, in the file {\bf test.f}, and  
then type "make test" and "./test.exe".
In \ref{tout}, the test outputs for $t$, $v$, and $\phi _k$ generation with
$P=1$, $E_b^{\rm Lab}$=4 GeV, $\theta_{\eta}=48^0$, 
20 bins for the histogramming and $10^8$ radiative events
are presented.

After generating all photonic variables for one radiative event, ELRADGEN 
reconstructs the four-momenta of the final particles. To make sure that the vectors are 
constructed properly, the next test corresponding
to $itest:=4$ is implemented. This test allows to perform the numerical
comparison of the generated variables $t$, $v$, and $\phi _k$ with the value
of these variables reconstructed from four-momenta of the particles. This test also reconstructs the mass of the real
photon that has to be equal to zero. 

The test with $itest:=5$ provides us with the comparison of the 
unpolarized cross section integrated over $v$ analytically and numerically.

%\vspace*{120mm}

%\section{Numerical comparison with elastic radiative tail from the BLAST data}  

\subsection{$v_{min}$-dependence and comparison with MASCARAD}
\label{check}
The Monte Carlo generator ELRADGEN 2.0 was developed
on the basis of the FORTRAN code MASCARAD, therefore the agreement
of outputs of both programs with the same input parameters has to be demonstrated as a primary test. 
Here we restrict our crosscheck to the JLab kinematic conditions
without cuts on inelasticity 
$v$ and focus on the comparison
of the ratio of the radiatively corrected cross section to the 
Born one. Define components of the cross sections as: 
\ba
\sigma^{L,T}_a(\xi_L,\eta_{L,T})=\sigma^u_a+P_LP_N\sigma^p_a(\xi_L,\eta_{L,T}),
\ea
where $a=0,BSV, rad, obs$

\begin{table*}
\label{tab:0}
\centering{
\begin{tabular}{|c|c|c|c|c|c|c|c|}
\hline
$v_{min}$ 
&
\multicolumn{2}{c|}{$\sigma^u _{rad}/\sigma^u_0$}
&
\multicolumn{2}{c|}{$\sigma^u _{BSV}/\sigma^u_0$}  
&
\multicolumn{3}{c|}{$weight=\sigma^u _{obs}/\sigma^u_0$}   
\\\cline{2-8}
GeV$^2$
&a
&b
&a
&b
&a
&b
&c
\\ 
\hline
1	              &1.144&1.145&0.9730&0.9737&2.117&2.119&\\ 
\cline{1-7}$10^{-1}$&1.316&1.317&0.8018&0.8018&2.118&2.118&\\ 
\cline{1-7}$10^{-2}$&1.478&1.473&0.6386&0.6386&2.116&2.111&2.117\\ 
\cline{1-7}$10^{-3}$&1.641&1.634&0.4754&0.4754&2.116&2.108&\\
\cline{1-7}$10^{-4}$&1.806&1.797&0.3122&0.3122&2.118&2.108&\\ \hline
\end{tabular}}
\vspace*{5mm}
\caption{The $v_{min}$-dependence of the ratios of radiative, BSV,
and observable contributions to the unpolarized ($P_LP_N\equiv 0$) electron-proton cross section to the Born cross section 
for JLab kinematic conditions ($E_{beam}=4$ GeV
and $Q^2=3$ GeV$^2$): a (b) presents the results of analytical (numerical) integration 
over $v$ in ELRADGEN, while c shows the results of the calculation using MASCARAD \cite{AAM}. 
}
\end{table*}
\begin{table*}
\centering{
\begin{tabular}{|c|c|c|c|c|c|c|c|c|}
\hline
$v_{min}$ 
&
\multicolumn{2}{c|}{$\sigma^L _{rad}/\sigma^L_0$}
&
\multicolumn{2}{c|}{$\sigma^L _{BSV}/\sigma^L_0$}  
&
\multicolumn{4}{c|}{$weight=\sigma^L _{obs}/\sigma^L_0$}   
\\\cline{2-9}
GeV$^2$
&
\multicolumn{2}{c|}{ELRADGEN}
&
\multicolumn{2}{c|}{ELRADGEN}  
&
\multicolumn{2}{c|}{ELRADGEN}   
&
\multicolumn{2}{c|}{MASCARAD}   
\\\hline
$P_LP_N$
&1
&-1
&1
&-1
&1
&-1
&1
&-1
\\ 
\hline
1	              &0.6277&1.275&0.9648&0.9765&1.592&2.251&&\\ 
\cline{1-7}$10^{-1}$&0.7913&1.448&0.8009&0.8080&1.592&2.250&&\\ 
\cline{1-7}$10^{-2}$&0.9469&1.605&0.6385&0.6487&1.585&2.243&1.591&2.249\\ 
\cline{1-7}$10^{-3}$& 1.106&1.764&0.4754&0.4893&1.582&2.240&&\\
\cline{1-7}$10^{-4}$& 1.269&1.927&0.3122&0.3300&1.582&2.240&&\\ \hline
\end{tabular}}
\vspace*{5mm}
\label{tab:1}
\caption{$v_{min}$-dependence of the radiative, BSV,
and observable contribution to electron-proton
scattering with longitudinally polarized target ($\theta _{\eta}=0$)
for a different spin orientation
in the Born units  and comparison with MASCARAD \cite{AAM} 
for JLab kinematic conditions ($E_{beam}=4$ GeV
and $Q^2=3$ GeV$^2$).
}
\end{table*}

\begin{table*}
\centering{
\begin{tabular}{|c|c|c|c|c|c|c|c|c|}
\hline
$v_{min}$ 
&
\multicolumn{2}{c|}{$\sigma^T _{rad}/\sigma^T_0$}
&
\multicolumn{2}{c|}{$\sigma^T _{BSV}/\sigma^T_0$}  
&
\multicolumn{4}{c|}{$weight=\sigma^L _{obs}/\sigma^T_0$}   
\\\cline{2-9}
GeV$^2$
&
\multicolumn{2}{c|}{ELRADGEN}
&
\multicolumn{2}{c|}{ELRADGEN}  
&
\multicolumn{2}{c|}{ELRADGEN}   
&
\multicolumn{2}{c|}{MASCARAD}   
\\\hline
$P_LP_N$
&1
&-1
&1
&-1
&1
&-1
&1
&-1
\\ 
\hline
1	              &0.9457&1.447&0.9730&0.9746&1.919&2.422&&\\ 
\cline{1-7}$10^{-1}$& 1.117&1.620&0.8018&0.8018&1.918&2.422&&\\ 
\cline{1-7}$10^{-2}$& 1.273&1.776&0.6386&0.6386&1.912&2.415&1.917&2.420\\ 
\cline{1-7}$10^{-3}$& 1.432&1.935&0.4754&0.4754&1.908&2.411&&\\
\cline{1-7}$10^{-4}$& 1.596&2.099&0.3122&0.3122&1.908&2.411&&\\ \hline
\end{tabular}}
\vspace*{5mm}
\label{tab:2}
\caption{$v_{min}$-dependence of the radiative, BSV
and observable contribution to  electron-proton
scattering with transversely polarized target ($\theta_{\eta}=\pi/2$, $\phi=\phi_{\eta}$) 
for different spin orientation
in the Born units and comparison with MASCARAD \cite{AAM} for JLab kinematic conditions 
($E_{beam}=4$ GeV
and $Q^2=3$ GeV$^2$). 
}
\end{table*}

First, we consider the unpolarized scattering for which an option with analytical integration over $v$ is available. Table 1 presents the results of 
 the analytical and numerical integration for the BVS- and radiative contributions ({\it i.e.}, $\sigma_{BSV}(v_{min})$ and $\sigma_{rad}(v_{min})$) to the observed cross section, as well as the results obtained using MASCARAD. Each of these contributions changes essentially by
decreasing $v_{min}$ from 1 to $10^{-4}$  GeV$^2$,  while 
the observable cross sections barely change  
for both analytical and numerical integration 
over $v$. The similar behavior 
of the radiative and BSV parts takes place for the
polarized case. This is illustrated in Tables 2 and 3. 
The observable cross sections change by no more than 1\%.

\subsection{Results from the BLAST data}  
\label{num}
% Tavi stuff:

Analyzing the $\Delta$-excitation region in $ep$-scattering,
it is necessary to extract the contribution of real hard-photon emission
that accompanies the elastic $ep$-scattering (so-called elastic radiative tail) 
and cannot be
removed  from the data by any experimental cuts. The main radiative photons  
are emitted by the electron leg (see Fig.\ref{fg} (d,e)), because their
contributions include the logarithm of the electron mass.  
These radiative events are spin-dependent, and
therefore affect not only the cross section, but other extracted quantities in 
the $\Delta$-region as well, {\it e.g. } asymmetries, spin-correlation parameters,
spin-structure functions, {\it etc}.

The BLAST experiment was designed to study spin-dependent electron scattering
off protons and deuterons with small systematic uncertainties \cite{Doug}.
The experiment used a longitudinally polarized, an intense electron beam and
isotropically pure highly-polarized internal targets of hydrogen and
deuterium from an atomic beam source.
For extraction of the elastic radiative tail contribution,  
the new version {\bf 2.0} of Monte Carlo generator ELRADGEN has been applied.
This generator was incorporated into the BLAST 
Monte Carlo event generator \cite{Tavi}, where longitudinally polarized electrons at an 
energy of 850 MeV and at a polarization factor of 65\%, were scattered off a 
highly-polarized hydrogen internal gas target ($P_N \sim  80$\%), with the 
average target spin direction oriented at $48.84^\circ$ to the left of the beam 
direction. 

\begin{figure}[t]
\vspace*{-7mm}
  \centering
  \includegraphics[width=14cm,height=8.55cm]{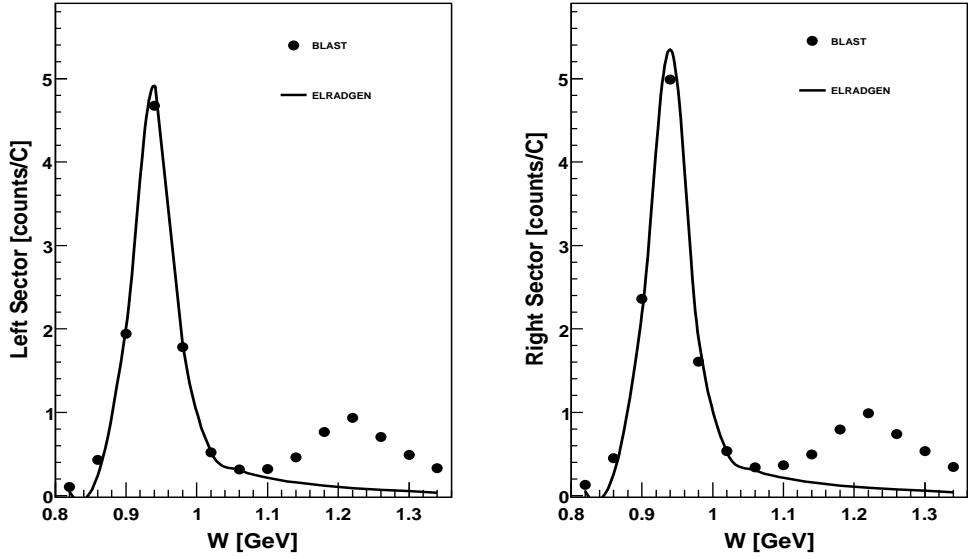}
  \caption{Normalized yields as a function of the invariant mass, 
    $W [{\rm GeV}]$ over $0.08 < Q^2 < 0.38 \, {\rm GeV}^2$.
    The dots show the BLAST ABS hydrogen data corrected for the background 
    contributions, and the solid line represents the Monte Carlo simulations 
    with radiative effects (ELRADGEN {\bf 2.0}).} 
  \label{fig:c_W_RC_Q2_08_38}
\vspace*{2mm}
\end{figure}
\vspace*{-2mm}
\begin{figure}[t!]
\vspace*{-2mm}
%\begin{figure}[h!]
  \centering
  \includegraphics[width=14cm,height=8.55cm]{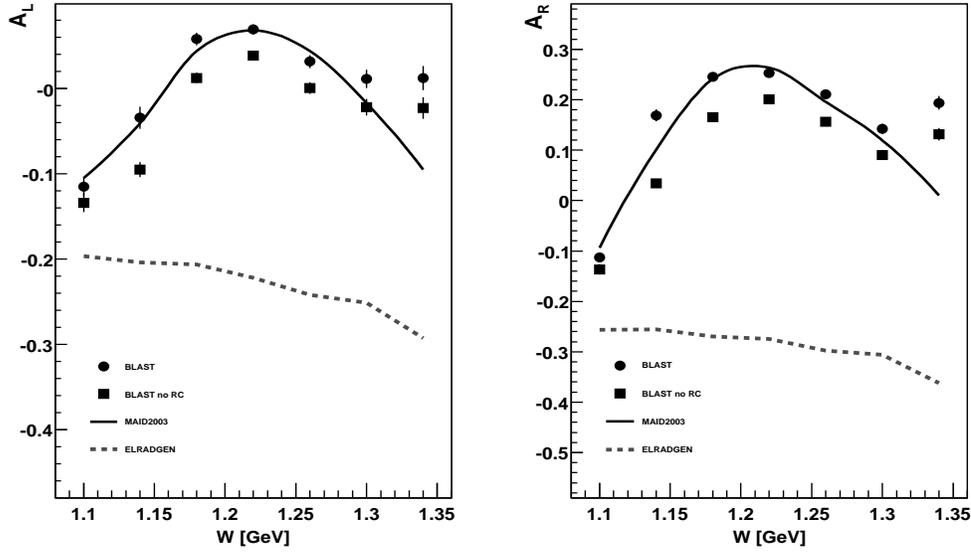}
%\vspace*{1mm}
  \caption{The effect of the radiative contributions to the asymmetry in 
    the $\Delta$-excitation region.
    The left (left) and right (right) asymmetries are shown with (dots)
    and without (squares) radiative corrections (RC), for 
    $0.08 < Q^2 < 0.38 \, {\rm GeV^2}$. 
    Monte Carlo simulations using the MAID 2003 model \cite{Dre99} 
    (straight line) and ELRADGEN (dotted line) 
    are shown for comparison. 
}
  \label{fig:c_A_W_Q2_08_38}
\end{figure}
\vspace*{1mm} 

In order to estimate the contribution of the elastic radiative tail to the 
$\Delta$-excitation region, the results of the Monte Carlo simulations 
were normalized to the data elastic peak, as shown in
Fig. \ref{fig:c_W_RC_Q2_08_38}. In this figure, the normalized yields
correspond to the outgoing electrons detected in each sector of the BLAST detector
(inclusive scattering).
The radiative tail obtained from the above normalization is subtracted 
from the measured yields (radiative corrections), and then the left and right 
asymmetries are extracted. For comparison, in Fig. \ref{fig:c_A_W_Q2_08_38} 
we show the left and right asymmetries with and without the radiative 
corrections. The asymmetry from ELRADGEN alone is also shown (dotted line) in
order to see the radiative tail effect to the overall asymmetries and its spin
dependence.

\section{Conclusion}  
\label{cconc}
In this paper, we presented a new version of Monte Carlo generator ELRADGEN for
simulation of real-photon events within
elastic lepton nucleon scattering 
for longitudinally polarized lepton and
arbitrary polarized target. Following the absolute necessity of both accuracy
and quickness for our program, we have developed the fast and highly
precise code using analytical integration wherever it was possible. The developed program has a broad spectrum of applications in data analysis of various experimental designs on polarized $ep$-scattering, including  the measurements of
 the generalized
parton distributions,  the generalized polarizabilities, and the evaluation of spin asymmetries in elastic scattering. Also, it can be used as a generator of the ``Born'' process in DVCS measurements and of the radiative tail from the elastic peak in DIS. The most significant application of the generator is in the experiments
with the complex detector geometry.

The set of numerical tests of the presented version of this code proved its high quality. First, a good agreement with FORTRAN code MASCARAD \cite{AAM} was found. Second, no dependence on the missing mass square
resolution was found. Third, the distributions of the generated radiative events are found to be
in accordance with the corresponding probability densities. Fourth, 
a good agreement with the radiative tail from the elastic peak 
 measured in the BLAST experiment was demonstrated.

Several additional steps allowing to make the simulation even faster are planned. They include implementation of 
analytical integration over variable $v$ of the 
hard-photon emission
contribution with the longitudinally
polarized target and 
utilizing the look-up table option for faster simulation
of radiative events with transverse component 
of the target polarization vector.

The spectrum of applications of the presented code could be extended after a certain substantive 
upgrade in several directions including the development of this generator for transferred
polarization from lepton beam
to recoil proton \cite{AhR}, and 
for involving this generator into the
measurement of the electromagnetic form-factors of the proton
in elastic scattering with 
unpolarized \cite{FFunp} and
polarized targets \cite{FFpol1,FFpol2}.
Inclusion of electroweak effects will provide the generalization for the investigation 
of electroweak corrections in experiments on axial
form factors of the nucleon \cite{Gor} and
parity violation elastic scattering \cite{Beck}. This generator can be included in data analysis of experiments
with the measurement of 
unpolarized and spin-flip
generalized polarizabilities
in virtual Compton scattering  
\cite{Gui,Dre98}.

  \section*{Acknowledgments}  
The authors would like to acknowledge useful discussion with
E.Tomasi-Gustafsson.
One of us (A.I.) would like to thank the staff
of MIT Bates Center for  their generous
hospitality during his visit.

\appendix
\renewcommand{\thesection}{Appendix~\Alph{section}}
\renewcommand{\thesubsection}{Appendix~\Alph{section}.\arabic{subsection}}
\renewcommand{\theequation}{\Alph{section}.\arabic{equation}}
\setcounter{equation}{0}
\def\thefigure{7}

\section{Four-vectors}
\label{fv}
\begin{figure}[t]
\unitlength 1mm
%\hspace*{-2cm}
%\vspace*{-5mm}
\begin{picture}(80,80)
\put(10,0){
\epsfxsize=12cm
\epsfysize=8cm
\epsfbox{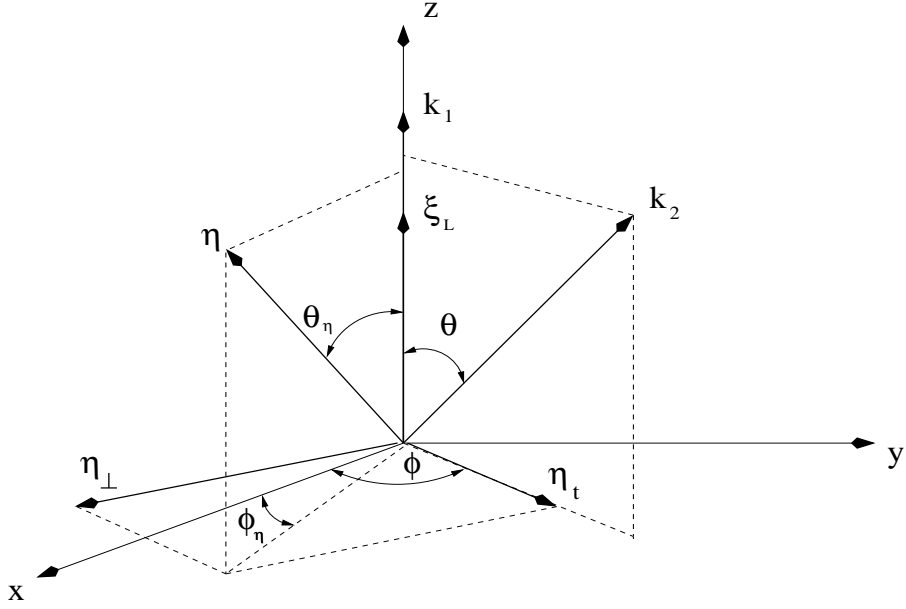}
}
\end{picture}
%\vspace*{9mm}
\caption{3-vectors decomposition in the LAB system}
\label{decv}
\end{figure}

In this section we present the explicit expression for
four-vectors decomposition in LAB system  depicted in Fig.~\ref{decv}

\subsection{Polarization vector definitions}
\label{pv}
As it was mentioned above we assume that
the electron beam has a longitudinal polarization. 
Therefore its polarization vector
has a form \cite{POLRAD}:
\ba
\xi_L=\frac 1{\sqrt{\lambda _s}}
\left (\frac Smk_1-2mp_1\right ).
\ea
The target polarization vector in the Lab. system
can be decomposed into  longitudinal 
\ba
\eta_L=\frac{1}{\sqrt{\lambda _s}}
\left (2Mk_1-\frac SMp_1\right )
\ea
and transverse $\eta_T$ components as it is depicted in Fig.~\ref{decv}
\ba
\eta=
\cos(\theta _{\eta })\eta _L+
\sin(\theta _{\eta })\eta _T,
\ea
where $\theta _{\eta }$ is the angle between 3-vectors 
${\bf k_1}$ and ${\boldsymbol \eta }$.
Transverse $\eta_T$ component can be presented as:
\ba
\eta_T=
\cos(\phi-\phi _{\eta })\eta _t+
\sin(\phi-\phi _{\eta })\eta _{\bot },
\ea
where $\phi _{\eta }$ is the angle  between (${\bf k_1}$, ${\boldsymbol \eta }$)
and OZX planes, and
\ba
\label{ett}
\eta_t&=&
\frac {(4m^2M^2+2 Q^2M^2-SX)k_1+\lambda _s k_2
-(SQ^2+2m^2S_x)p_1}{\sqrt{\lambda \lambda _s}},
\nonumber \\[2mm]
\eta_\bot&=&
\left(0,\frac{\bf{k_2}\times \bf{k_1}}{ |\bf{k_2}||\bf{k_1}|\sin \theta }
\right)=(0,\sin \phi,-\cos \phi,0).
\ea
Here 
\ba 
&X=2k_2p_1,\; S_x=S-X,\;
\lambda=S XQ^2-M^2Q^4-m^2\lambda_q,
\nonumber \\
&\lambda_q=S_x^2+4Q^2M^2.
\label{invr}
\ea
 
It should be noted, that
for the BSV process the variable $X$ is fixed by $Q^2$ and $S$:
$X=S-Q^2$, while for the radiative one, the variable $X$ (as well as $S_x$ and 
$\lambda_q $) depends on inelasticity:
\ba
X=S-Q^2-v,\;S_x=Q^2+v,\;\lambda_q=(Q^2+v)+4Q^2M^2.
\label{kr}
\ea

As it follows from (\ref{vvk0}) the normal to scattering plane
component of $\eta $ 
satisfies the equations: 
\ba
k_1 \eta _\bot=k_2 \eta_\bot=p_1 \eta_\bot=0,\;\;
%\nonumber \\
k \eta_\bot=-p_2 \eta_\bot=\sin \phi _k\frac{\sqrt{\lambda_3}}{\sqrt{\lambda_q}}.
\label{invr2}
\ea

\subsection{Four-momenta reconstruction}
\label{av}
After generation of photonic variable $t$, $v$ and $\phi _k$ the
four-momenta of final 
proton $\displaystyle p_2=(p_2^{(0)},p_2^{(1)},p_2^{(2)},p_2^{(3)})$,
lepton $\displaystyle k_2=(k_2^{(0)},k_2^{(1)},k_2^{(2)},k_2^{(3)})$
and real photon $\displaystyle k=(k^{(0)},k^{(1)},k^{(2)},k^{(3)})$
 in the LAB system read:
\begin{eqnarray}
&&p^{(1)}_2=\frac{\sqrt{\lambda _3}
(
\lambda _1\cos \phi \cos \phi _k
-
\sqrt{\lambda _q}S\sin \phi \sin \phi _k
)
+\lambda _2\sqrt{\lambda _4}\cos \phi
}{\lambda _q S},\nonumber \\
&&p^{(2)}_2=\frac{\sqrt{\lambda _3}
(
\sqrt{\lambda _q}S\cos \phi \sin \phi _k
+\lambda _1\sin \phi \cos \phi _k
)
+\lambda _2\sqrt{\lambda _4}\sin \phi
}{\lambda _q S},\nonumber \\
&&p^{(3)}_2
=\frac{\lambda _1 \lambda _2-4M^2
\sqrt{\lambda _3 \lambda _4}\cos \phi _k
}{2\lambda _q M S},\; 
%\nonumber \\&&
p_2^{(0)}=\frac{t+2M^2}{2M} ,\nonumber \\
&&
k_2^{(1)}=\frac{\sqrt{\lambda _4}\cos \phi }{S},\;
%\nonumber \\&&
k_2^{(2)}=\frac{\sqrt{\lambda _4}\sin \phi }{S},\;
%\nonumber \\&&
k_2^{(3)}=\frac{S^2-\lambda _1}{2MS},\;
%\nonumber \\&&
k_2^{(0)}=\frac{S-Q^2-v}{2M},
\nonumber \\
&&k^{(1)}=\frac{\sqrt{\lambda _3}
(
\sqrt{\lambda _q}S\sin \phi \sin \phi _k
-
\lambda _1\cos \phi \cos \phi _k
)
+(\lambda _q-\lambda _2)\sqrt{\lambda _4}\cos \phi
}{\lambda _q S},\nonumber \\
&&
k^{(2)}=\frac{-\sqrt{\lambda _3}
(
\sqrt{\lambda _q}S\cos \phi \sin \phi _k
+\lambda _1\sin \phi \cos \phi _k
)
+(\lambda _q-\lambda _2)\sqrt{\lambda _4}\sin \phi
}{\lambda _q S},
\nonumber 
\\&&
k^{(3)}
=\frac{\lambda _1 (\lambda _q-\lambda _2 )+4M^2
\sqrt{\lambda _3 \lambda _4}\cos \phi _k
}{2\lambda _q M S},\;
%\nonumber \\&&
k^{(0)}=\frac{Q^2+v-t}{2M},
\label{vvk0}
\end{eqnarray}
where
\begin{eqnarray}
\lambda _1=S(Q^2+v)+2M^2Q^2,\;
\qquad
\qquad
&&\lambda _2=t(Q^2+v)+2M^2(Q^2+t),
\nonumber \\[2mm]
\lambda _3=tv(Q^2-t+v)-M^2(Q^2-t)^2,
\;
&&\lambda _4=Q^2S(v_{max}-v).
\label{vvk}
\end{eqnarray}

\section{Explicit expressions for the kinematic quantities $\theta $}
\label{exp}
%\subsection{Kinematic coefficients $\theta _i^B$ and  $\theta _i(v_1,v_2)$}
The kinematic coefficients $\theta _i^B$
appear as a convolution of the leptonic tensor
$L_{\mu \nu}^B$
with corresponding hadronic structures 
\ba
&&w^{\mu\nu}_1=-g_{\mu\nu}, 
\qquad
\qquad
\qquad
\;\;
w^{\mu\nu}_2=\frac{p_{\mu}p_{\nu}}{M^2},
\nonumber \\
&&w^{\mu\nu}_3=-iP_N\epsilon_{\mu\nu\lambda\sigma}{ q_{\lambda}\eta_{\sigma}
\over M}, \qquad
w^{\mu\nu}_4=iP_N\epsilon_{\mu\nu\lambda\sigma}{ q_{\lambda}p_{\sigma}\;
\eta q \over M^3}
\label{hadst}
\ea
and read
\ba
&&\theta_1^B=\frac 12 L^B_{\mu \nu }w_1^{\mu \nu }= Q^2,
\nonumber \\ 
&&\theta_2^B=\frac 12 L^B_{\mu \nu }w_2^{\mu \nu }=\frac 1 {2M^2}(S(S-Q^2)-M^2Q^2),
\nonumber \\ 
&&\theta_3^B=\frac 12 L^B_{\mu \nu }w_3^{\mu \nu }=P_LP_N\frac{2m}{M}(q\eta \; k_2\xi-\xi \eta \;Q^2),
\nonumber \\ 
&&\theta_4^B=\frac 12 L^B_{\mu \nu }w_4^{\mu \nu }=P_LP_N\frac {mQ^2\; q\eta}  {M^3}(2p_1\xi -k_2\xi ).
\label{thb}
\ea
Here $P_L$ and $P_N$ define the degree of the lepton and target polarization  respectively and 
the explicit expressions for polarized vectors of the scattering particles $\xi$ and $\eta $
can be found in \ref{pv}.

The quantities $\theta _i (v_1,v_2)$ appear as a convolution of the leptonic
tensor that is responsible for the real photon emission:
\ba
L_{\mu \nu}^R
&=&-\frac 12 {\rm Tr}[({\hat k}_2+m)\Gamma _{\mu \alpha }
(1-P_L{\hat \xi}\gamma _5 )
({\hat k}_1+m)
{\hat \Gamma }_{\alpha \nu }],
\nonumber \\
\Gamma _{\mu \alpha }&=&
\left (
\frac {k_{1\alpha }}{k k_1}-\frac {k_{2\alpha }}{k k_2}\right )\gamma_{\mu}
-\frac{\gamma _{\mu} {\hat k}\gamma _{\alpha}}{2kk_1}
-\frac{\gamma _{\alpha} {\hat k}\gamma _{\mu}}{2kk_2},\;
\nonumber \\
\hat {\Gamma }_{\alpha \nu }&=&
\left (
\frac {k_{1\alpha }}{k k_1}-\frac {k_{2\alpha }}{k k_2}\right )\gamma_{\nu}
-\frac{\gamma _{\alpha} {\hat k}\gamma _{\nu}}{2kk_1}
-\frac{\gamma _{\nu} {\hat k}\gamma _{\alpha}}{2kk_2}
\ea
with hadronic structures presented in eq.(\ref{hadst}).
 
As a result
\ba
\theta _i^R (v_1,v_2)
&=&-\frac 1 {4\pi }\int \limits _{v_1}^{v_2}\frac {dv}{\sqrt{\lambda _q}}
\int \limits _0^{2\pi}d\phi _k \
L^R_{\mu \nu }w_i^{\mu \nu }(q\to q-k )=
\sum_{j=1}^{k_i}
\int \limits _{v_1}^{v_2}dv
R^{j-3}\theta_{ij}^R (v )
\nonumber \\
&=&
\sum_{j=1}^{k_i}
\int \limits _{v_1}^{v_2}dv
\int \limits _0^{2\pi}d\phi _k \
R^{j-3}\theta_{ij}^R (v, \phi _k ),\;
\label{thr}
\ea
where $R$ and $k_i$ are defined after eq. (\ref{thr000}).

\subsection{Quantities $\theta_{ij} (v )$ and $\theta_{ij} (v, \phi _k )$}
\label{exp1}
Here, we combine 
explicit expressions for 
$\theta_{ij}^R(v)$ and $\theta_{ij}^R(v,\phi _k)$ quantities
calculated for polarized scattering \cite{ASh,POLRAD,HAPRAD} and present the explicit 
expressions $\theta _i^R (v_1,v_2) $ calculated 
for unpolarized scattering only.

Both types of quantities 
$\theta_{ij}^R(v)$ and
$\theta_{ij}^R(v,\phi _k)$ 
for $i=1,2,3$ take the similar form 
%\newpage
\begin{eqnarray}
\theta _{11}^R&=& 4Q^2F_{IR},
\nonumber\\[0.3cm]
\theta _{12}^R&=& 4\ta F_{IR},
\nonumber\\[0.3cm]
\theta _{13}^R&=& -4F - 2\ta^{2}F_{d},
\nonumber\\[0.3cm]
\theta _{21}^R&=& {2(SX - M^{2}Q^2)F_{IR}/M^{2}},
\nonumber\\[0.3cm]
\theta _{22}^R&=&(
2m^{2}S_{p}F_{2-}+S_{p}S_{x}F_{1+}+2(S_{x}-2M^{2}\ta)F_{IR}-\ta S
 ^{2}_{p}F_{d}
)
/2M^{2},
\nonumber\\[0.3cm]
\theta _{23}^R&=&
(4M^{2}F+(2M^{2}\ta-S_{x})\ta F_{d}-S_{p}F_{1+})
/2M^{2},
\nonumber\\[0.3cm]
\theta _{31}^R&=&P_LP_N{8m\over M}(\eta q\; k_{2}\xi
- Q^2\; \xi \eta )F_{IR},
\nonumber\\[0.3cm]
\theta _{32}^R&=&-P_LP_N{2m\over M}
(2\eta q\; (\ta k_{2}\xi F_d-2F^{\xi }_{IR})
+Q^2\eta {\mathcal K}(F^{\xi }_{2+}-F^{\xi }_{-2}-2F^{\xi }_d)
\nonumber\\[0.3cm]&&
+4\xi \eta \ta F_{IR}-4 m^2k_{2}\xi(2 F^{\eta }_d-F^{\eta }_{2+})) 
\nonumber\\[0.3cm]
\theta _{33}^R&=&P_LP_N{2m\over M}(\eta {\mathcal  K}\;\ta 
(2F^{\xi }_{d}+F^{\xi }_{2-}-F^{\xi }_{2+})
-2\;k_{2}\xi\;  \ta F^{\eta }_{d} 
- 4m^{2}F^{\xi \eta }_{d}
- 6F^{\xi \eta }_{IR} 
 \nonumber\\[0.3cm]&&
+ 
Q^2(F^{\xi \eta
 }_{2+}
- 
F^{\xi \eta }_{2-} 
)),
\nonumber\\[0.3cm]
\theta _{34}^R&=&P_LP_N{2m\ta \over M}\pmatrix{2F^{\xi \eta
 }_{d}+F^{\xi \eta }_{2+}-F^{\xi \eta }_{2-}},
\label{th13}
\end{eqnarray}
where $\ta =(t-Q^2)/R$ and the four-vector ${\cal K}=k_1+k_2$.

For $i=4$ we have
\begin{eqnarray}
\theta_{41}^R=\eta q{\widetilde \theta_{41}}/M,
\qquad 
\qquad 
\;\;
&
%\nonumber\\[0.3cm]
\theta_{42}^R=(\eta q{\widetilde \theta}_{42}-\widetilde \theta_{41}^{\eta })/M,
\nonumber\\[0.3cm]
\theta_{43}^R=(\eta q{\widetilde \theta}_{43}-\widetilde \theta_{42}^{\eta })/M,
\;\;\;\;\;\; 
&
%\nonumber\\[0.3cm]
\theta_{44}^R=(\eta q{\widetilde \theta}_{44}-\widetilde \theta_{43}^{\eta })/M,
\nonumber\\[0.3cm]
\theta_{45}^R=-\widetilde \theta_{44}^{\eta }/M,
\qquad 
\qquad 
\;\;\;\;
&
\label{th41}
\end{eqnarray}
where
\begin{eqnarray}
{\widetilde \theta}_{41}&=&P_LP_N{4m\over M^{2}}
\pmatrix{2\;\xi p\; Q^2-S_{x}\;\xi k_{2}}F_{IR},
\nonumber\\[0.3cm]
{\widetilde \theta} _{42}&=&P_LP_N{m\over M^{2}}(
2\pmatrix{S_{p}-2S_{x}}F^{\xi }_{IR}
+ 2\;k_{2}\xi\;
\ta S_{x}F_{d}
+ 8\;\xi p\; \ta F_{IR}
\nonumber\\[0.3cm]
&&+
 S_{p}\pmatrix{Q^2_{m}F^{\xi }_{2+}-Q^2F^{\xi }_{2-}}
-
4m^{2}
 (\;k_{2}\xi\; (2F_{d}- F_{2+})
+S_{p}(F^{\xi }_{2+}
-F^{\xi }_{d})
),
\nonumber\\[0.3cm]
{\widetilde \theta }_{43}&=&P_LP_N
{m\over M^{2}} (
\pmatrix{Q^2-\ta S_{p}}F^{\xi }_{2-}
- \pmatrix{Q^2_{m}-\ta S_{p}}F^{\xi }_{2+}
+ 2\;k_{2}\xi\;  \ta F_{d}
\nonumber
\\[0.3cm]
&& 
+ 6F^{\xi}_{IR}
- 2F^{\xi }_{d}\ta S_{p}
),
\nonumber\\[0.3cm]
{\widetilde \theta} _{44}&=&-P_LP_N{m\ta\over M^{2}}\pmatrix{2F^{\xi
}_{d}-F^{\xi
 }_{2-}+F^{\xi }_{2+}}.
\label{th42}
\end{eqnarray}
The quantities ${\widetilde \theta} _{4j}^{\eta }$
are calculated as:
\ba
{\widetilde \theta} _{4j}^{\eta }=
{\widetilde \theta} _{4j}
(F_{all}\to F_{all}^{\eta}, F_{all}^{\xi}\to F_{all}^{\xi \eta}).
\ea
The upper indices in $F_{all}$ appears in the following way: 
\ba
2F_{2+}^{\xi}&=&
(2F_{1+}+\ta F_{2-})s_{\xi}+F_{2+}r_{\xi},
\nonumber\\[0.3cm]
2F_{2+}^{\eta }&=&
(2F_{1+}+\ta F_{2-})s_{\eta }+F_{2+}r_{\eta}+
\frac{4\sin \phi_k d_{\eta}}R\sqrt{\frac{\lambda _3}{\lambda _q}}F_{2+},
\nonumber\\[0.3cm]
2F_{2-}^{\xi}&=&(2
F_{d}+F_{2+})\ta s_{\xi}+F_{2-}r_{\xi},
\nonumber\\[0.3cm]
2F_{d}^{\xi}&=&F_{1+}s_{\xi}+F_{d}r_{\xi},
\nonumber\\[0.3cm]
2F_{d}^{\eta}&=&F_{1+}s_{\eta}+F_{d}r_{\eta}+
\frac{4\sin \phi_k d_{\eta}}R\sqrt{\frac{\lambda _3}{\lambda _q}}F_{d},
\nonumber\\[0.3cm]
4F_{2+}^{\xi\eta}&=&
(2F_{1+}+\ta F_{2-})
(r_{\eta}s_{\xi}+s_{\eta}r_{\xi})
+F_{2+}
(r_{\eta}r_{\xi }+\ta^2 s_{\eta}s_{\xi})
\nonumber\\[0.3cm]
&&+4(2F+F_d\ta^2)s_{\eta}s_{\xi}
+\frac{8\sin \phi_k d_{\eta}}R\sqrt{\frac{\lambda _3}{\lambda _q}}F_{2+}^{\xi },
\nonumber\\[0.3cm]
4F_{2-}^{\xi\eta}&=&
(2F_{d}+ F_{2+})
(r_{\eta}s_{\xi}+s_{\eta}r_{\xi})
+F_{2-}
(r_{\eta}r_{\xi}+\ta^2 s_{\eta}s_{\xi})
+4\ta F_{1+}s_{\eta}s_{\xi}
\nonumber\\[0.3cm]
&&
+\frac{8\sin \phi_k d_{\eta}}R\sqrt{\frac{\lambda _3}{\lambda _q}}F_{2-}^{\xi },
\nonumber
\\[0.3cm]
4F_{d}^{\xi\eta}&=&
F_{1+}
(r_{\eta}s_{\xi}+s_{\eta}r_{\xi})
+F_{d}
(r_{\eta}r_{\xi}+\ta^2 s_{\eta}s_{\xi})
+4 Fs_{\eta}s_{\xi}
\nonumber
\\[0.3cm]
&&+\frac{8\sin \phi_k d_{\eta}}R\sqrt{\frac{\lambda _3}{\lambda _q}}F_{d}^{\xi }.
\end{eqnarray}
The quantities
\ba
s_{\{\xi,\eta\}}=a_{\{\xi,\eta\}}+b_{\{\xi,\eta\}},
r_{\{\xi,\eta\}}=\ta (a_{\{\xi,\eta\}}-b_{\{\xi,\eta\}})
+2c_{\{\xi,\eta\}}
\label{th10}
\ea
are combinations of coefficients of polarization vectors
$\xi$ and
$\eta$ expansion over basis (see \ref{pv})
\ba
\xi&=&2(
a_{\xi} k_1 +
b_{\xi} k_2 +
c_{\xi} p),
\nonumber
\\
\eta&=&2(
a_{\eta} k_1 +
b_{\eta} k_2 +
c_{\eta} p+d_{\eta}\eta_\bot ).
\label{th11}
\ea
We note that the scalar products from (\ref{th13},\ref{th41},\ref{th42}) are also
calculated in  terms of the polarization vector coefficients:
\ba
&\eta q =-Q^2(a_{\eta}-b_{\eta})+S_xc_{\eta},   \quad
\eta {\cal K} =(Q^2+4m^2)(a_{\eta}+b_{\eta})+S_pc_{\eta},   
\nonumber \\[2mm]&
k_2\xi =Q^2_ma_{\xi}+2m^2b_{\xi}+Xc_{\xi},   \quad
\xi p =Sa_{\xi}+Xb_{\xi}+2M^2c_{\xi},   
\nonumber \\&
\frac 12 \xi\eta=
2m^2(a_{\xi}a_{\eta}+b_{\xi}b_{\eta})
+2M^2c_{\xi}c_{\eta}
+Q_m^2(a_{\xi}b_{\eta}+b_{\xi}a_{\eta})
\nonumber \\&
+S(a_{\xi}c_{\eta}+c_{\xi}a_{\eta})
+X(b_{\xi}c_{\eta}+c_{\xi}b_{\eta}).
\label{th16}
\ea

With the exception of the contribution proportional to $d_{\eta}$
all dependencies of $\theta _{ij}$ on the photonic variable $\phi _k $
are included in the quantities $F$, however in both cases we have:
\ba
F_{IR}=m^2F_{2+}-Q^2F_d
\label{fir00}
\ea
%All formulae (B1-B2) of ref.\cite{ASh} or (B.1-B.11) of ref.\cite{POLRAD} can be retained for our case. Instead of functions
%$F's$ from (B5) \cite{ASh} or (B.12) \cite{POLRAD} we use the following expressions

So for $\theta _{ij}(v,\phi _k) $ the quantities $F$ read 
\ba
&&F_{d}(v,\phi _k)=\frac{F(v,\phi _k)}{z_1z_2},
\qquad
\qquad
\;\;\;\;
F_{1+}(v,\phi _k)=F(v,\phi _k)\left (\frac 1{z_1}+\frac 1{z_2}\right ),\nonumber \\
&&F_{2\pm}(v,\phi _k)=F(v,\phi _k)\left(\frac 1{z_2^2}\pm\frac{1}{z_1^2}\right ),
\; 
F(v,\phi _k)=\frac 1{2\pi\sqrt{\lambda_q}}.
\label{fir001}
\ea
Here 
\begin{eqnarray}
z_1&=&\frac{2kk_1}R=\frac{1}{\lambda_q}(Q^2S_p+\ta(SS_x+2M^2Q^2)-2M\sqrt{\lambda_z}\cos\phi_{k}),
\nonumber \\
z_2&=&\frac{2kk_2}R=\frac{1}{\lambda_q}(Q^2S_p+\ta(XS_x-2M^2Q^2)-2M\sqrt{\lambda_z}\cos\phi_{k}),
\end{eqnarray}
and
\ba
&&\lambda_z
=(\ta-\ta_{min})(\ta_{max}-\ta)\lambda,\qquad 
\ta _{max/min}=\frac{S_x\pm\sqrt{\lambda _q}}{2M^2},
\nonumber \\
&&S_p=S+X=2S-Q^2-v.
\ea

The following equalities define the functions $F$
for $\theta _{ij}^R(v) $:
\begin{eqnarray}
&&F(v) =   \lambda ^{-1/2}_{q},
\;
F_{d} (v)=   \ta ^{-1} (C^{-1/2}_{2}(\ta )-C^{-1/2}_{1}(\ta )),
\nonumber\\[2mm]
&&
F_{1+} (v)=  C^{-1/2}_{2}(\ta ) + C^{-1/2}_{1}(\ta ),
\nonumber\\[2mm]
&&
F_{2\pm} (v)=B_2(\ta ) C^{-3/2}_2(\ta )\mp B_1(\ta )C^{-3/2}_1(\ta ),
\label{ffv}
\end{eqnarray}
 where
\ba
&&B_{1,2}(\ta )= - {1\over 2} \pmatrix{\lambda _{q}\ta \pm
 S_{p}(S_{x}\ta +2Q^2)},
\nonumber \\[0.2cm]
&&C_{1}(\ta ) =(S\ta + Q^2)^{2}+ 4m^{2}(Q^2 + \ta S_x - \ta^{2}M^{2}),
\nonumber  \\[0.2cm]
&&C_{2}(\ta ) =(X\ta - Q^2)^{2}+ 4m^{2}(Q^2 + \ta S_x - \ta^{2}M^{2}).
\label{bc}
\ea

We note that $F_d$ has a $0/0$-like uncertainty  for $\ta =0$
(inside the integration region). It leads to difficulties in
numerical integration, so another form is used also
\begin{equation}
F_d (v)={S_p(\ta S_x+2Q^2)\over {C_1^{1/2}(\ta )}{C_2^{1/2}(\ta )}
({C_1^{1/2}(\ta )}+{C^{1/2}_2(\ta )})}.
\end {equation}

\subsection{Quantities $\theta_{i}^R (v_1,v_2)$}
\label{expth}
As it was mentioned above for the unpolarized scattering, the integration over $v$ is
performed analytically resulting in:  
\begin{eqnarray}
\theta _1^R (v_1,v_2)&=&-4I_F
+4tI^{-2}_{2+}-2(Q^4+t^2)I^{-2}_{d},
\nonumber \\
\theta _2^R (v_1,v_2)&=&\frac 1{2M^2}(4M^2I_F
-t(I_{1+}^0+2I_d^{0})
+t(2S-t)I_{1+}^{-1}
+4I_{21}^0
\nonumber \\&&
-4(2S-t)I_{21}^{-1}
+4(S^2-t(M^2+S))I_{2+}^{-2}
\nonumber \\&&
+t(Q^2+4S-3t)I_d^{-1}
\nonumber \\&&
+[t(Q^2t-(2S-t)^2)+2M^2(t^2+Q^4)]I_d^{-2}),
\end{eqnarray}
where
\begin{eqnarray}
\label{II}
I_F&=&\int \limits^{v_2}_{v_1} dv F (v)=
\log\left(\frac
{Q^2+v_2+\sqrt{(Q^2+v_2)^2+4M^2Q^2}}
{Q^2+v_1+\sqrt{(Q^2+v_1)^2+4M^2Q^2}}
\right),
\nonumber \\
I_{1+}^0&=&\int \limits^{v_2}_{v_1} dv F_{1+} (v)=(Q^2-t)\left(
\frac S{Q^4}\Delta L_1-
\frac {S-t}{t^2}\Delta L_2\right)
+\frac{\Delta_1^1}{Q^4}
+\frac{\Delta_2^1}{t^2},
\nonumber \\
I_{1+}^{-1}&=&\int \limits^{v_2}_{v_1}\frac{dv}R F_{1+} (v)=
\frac 1{Q^2}\Delta L_1+
\frac 1t\Delta L_2,
\nonumber \\
I_{21}^0&=&\frac 1 2m^2\int \limits^{v_2}_{v_1}dv(F_{2+}(v)-F_{2-}(v))=\frac12 (Q^2-t)
\frac{S^2}{Q^4}
\Delta _1^0,
\nonumber \\
I_{21}^{-1}&=&\frac 12 m^2\int \limits^{v_2}_{v_1}\frac {dv}R(F_{2+}(v)-F_{2-}(v))=
\frac{S}{2Q^2}
\Delta _1^0,
\nonumber \\
I_{2+}^{-2}&=&m^2\int \frac{dv}{R^2}F_{2+} (v)
=\frac 1{2(Q^2-t)}
\left(
\Delta _1^0
-
\frac{Q^2}{t}
\Delta _2^0
\right),
\nonumber \\
I_{d}^{0}&=&\int \limits^{v_2}_{v_1}dvF_d (v)=
(Q^2-t)\left(\frac {S^2}{Q^6}\Delta L_1-\frac{(S-t)^2}{t^3}\Delta L_2
\right)
+2\frac S{Q^6}\Delta _1^1
\nonumber \\&&
+2\frac {S-t}{t^3}\Delta _2^1
+\frac 1{2(Q^2-t)}\left(
\frac {\Delta _1^2}{Q^6}-
\frac {\Delta _2^2}{t^3}
\right),
\nonumber \\
I_{d}^{-1}&=&\int \limits^{v_2}_{v_1}\frac{dv}RF_d (v)=
\frac S{Q^4}\Delta L_1+\frac {S-t}{t^2}\Delta L_2
+\frac 1{Q^2-t}\left(
\frac {\Delta _1^1}{Q^4}-
\frac {\Delta _2^1}{t^2}
\right),
\nonumber \\
I_{d}^{-2}&=&\int \limits^{v_2}_{v_1}\frac{dv}{R^2}F_d (v)=
\frac 1{Q^2-t}
\left(\frac 1{Q^2}\Delta L_1
-\frac 1t\Delta L_2
\right).
\end{eqnarray}
Here
\ba
&&\Delta_i^2=|D_i(v_2)|D_i(v_2)-|D_i(v_1)|D_i(v_1),
%\nonumber \\
\;
\Delta_i^1=|D_i(v_2)|-|D_i(v_1)|,
\nonumber \\[2mm]
&&\Delta_i^0=D_i(v_2)/D^2_{i+3}(v_2)
-D_i(v_1)/D^2_{i+3}(v_1),
 \\[2mm]
&&\Delta L_1=
\log \Biggl[
\frac
{2m^2t(Q^2-t+2v_2)+Q^2D_1(v_2)+D_4(v_2)\sqrt{4m^2t+Q^4}}
{2m^2t(Q^2-t+2v_1)+Q^2D_1(v_1)+D_4(v_1)\sqrt{4m^2t+Q^4}}
\Biggr],
\nonumber \\
&&\Delta L_2=
\log \Biggl[
\frac
{2m^2\sqrt{t}(Q^2-t+2v_2)+\sqrt{t}D_2(v_2)+D_5(v_2)\sqrt{4m^2+t}}
{2m^2\sqrt{t}(Q^2-t+2v_1)+\sqrt{t}D_2(v_1)+D_5(v_1)\sqrt{4m^2+t}}
\Biggr]
\nonumber
\ea
and
\begin{eqnarray}
D_1(v)&=&(t-Q^2)(S-Q^2)+Q^2v,
\nonumber \\
D_2(v)&=&S(Q^2-t)+tv,
\nonumber \\
D_3(v)&=&vt(v-t+Q^2)-M^2(Q^2-t)^2,
\nonumber \\
D_4(v)&=&\sqrt{D^2_1(v)+4 m^2D_3(v)},
\nonumber \\
D_5(v)&=&\sqrt{D^2_2(v)+4 m^2D_3(v)}.
\end{eqnarray}

\section{Test output}
\label{tout}
Here, we present the results of the test as {\bf test.dat} output file
corresponding to:\\
1) $itest:=1$ -- the generation of $\rho(t)$ distribution 
and comparison with the analytical cross section
corresponding to the first formula in (\ref{roo})
(here and below invariants $v, t$ are in $\mbox{GeV}^2$)
\input{testt.dat}
2) $itest:=2$ -- the generation of $\rho(v)$ distribution 
and comparison with the analytical cross section
corresponding to the second formula in (\ref{roo})
\input{testv.dat}
3) $itest:=3$ -- the generation of $\rho(\phi _k)$ distribution 
and comparison with the analytical cross section
corresponding to the third formula in (\ref{roo})
\input{testphi.dat}
4) $itest:=4$ -- the cross-check 
of the accuracy of the vector reconstruction for 5 random radiative events 
\input{testvect.dat}
4) $itest:=5$ -- the cross-check 
of the accuracy of the analytical and numerical integration over $t$  
for unpolarized scattering
\input{testint.dat}


\begin{thebibliography}{999}
\bibitem{Diehl}
  M.~Diehl,
  %``Generalized parton distributions,''
  Phys.\ Rept.\  388, (2003) 41. 
\bibitem{HERMES}
A.~Airapetian {\it et al.} % [HERMES Collaboration],
  %``Measurement of azimuthal asymmetries associated with deeply virtual Compton
  %scattering on an unpolarized deuterium target,''
  Nucl.\ Phys.\  B 829,  (2010) 1.
\bibitem{Gui}
P. A. M. Guichon \emph{et al.},  Nucl.\ Phys.\ A 591, (1995) 606.
\bibitem{Dre98}
D. Drechsel \emph{et al.}, Phys.\ Rev.\ C 57, (1998) 941.
\bibitem{ASh}
I.~V.~Akushevich and N.~M.~Shumeiko,
J.\ Phys.\ G 20 (1994) 513.
\bibitem{ABKR}
A. Akhundov, D. Bardin, L. Kalinovskaya 
and T. Riemann, Fortsch. Phys. 44, (1996)  373.
\bibitem{MaxTj}
L.~C.~Maximon and J.~A.~Tjon,
%``Radiative corrections to electron proton scattering,''
Phys.\ Rev.\ C 62,  (2000) 054320.

\bibitem{MoTsai}
L.~W.~Mo and Y.~S.~Tsai,
%``Radiative Corrections To Elastic And Inelastic E P And Mu P Scattering,''
Rev.\ Mod.\ Phys.\  41, (1969) 205.

\bibitem {Ent}
R. Ent  \emph{et al.}
Phys. Rev. C {\bf 64} (2001) 054610.
\bibitem {BSh}
D. Yu. Bardin,  N.M. Shumeiko:
Nucl. Phys. B 127 (1977) 242.
%\bibitem{AABJ}
%A.V. Afanasev, I. Akushevich, V. Burkert, K. Joo:
%Phys.Rev. D {\bf 66} (2002) 074004.


\bibitem{AAM}
A. V. Afanasev, I. Akushevich, N. P. Merenkov:
Phys. Rev. D 64 (2001) 113009.
\bibitem {AAIM}
A. V. Afanasev, I. Akushevich, A. Ilyichev, N. P. Merenkov:
Phys. Lett. B 514 (2001) 269.
\bibitem{JLep1}
  R.~Madey {\it et al.}  [E93-038 Collaboration],
  %``Measurements of G(E)(n)/G(M)(n) from the H-2(e(pol.),e' n(pol.))H-1
  %reaction to Q**2 = 1.45-(GeV/c)**2,''
  Phys.\ Rev.\ Lett.\  {\bf 91}, 122002 (2003)
  [arXiv:nucl-ex/0308007].
  %%CITATION = PRLTA,91,122002;%%
\bibitem{JLep2}
D.~I.~Glazier {\it et al.},
  %``Measurement of the Electric Form Factor of the Neutron at Q^2 = 0.3-0.8
  %(GeV/c)^2,''
  Eur.\ Phys.\ J.\  A {\bf 24}, 101 (2005)
  [arXiv:nucl-ex/0410026].
  %%CITATION = EPHJA,A24,101;%%

\bibitem{ESF}
E.~A.~Kuraev, N.~P.~Merenkov and V.~S.~Fadin,
%``Calculation Of The Radiative Corrections To The Cross Section For  Electron Nucleus Scattering By The Method Of Structure Functions,''
Sov.\ J.\ Nucl.\ Phys.\  47, (1988) 1009.
\bibitem{ESF1}
  A.~V.~Afanasev, I.~Akushevich and N.~P.~Merenkov,
  %``QED correction to asymmetry for polarized e p scattering from the  method
  %of the electron structure functions,''
  J.\ Exp.\ Theor.\ Phys.\  {\bf 98}, 403 (2004)
  [Zh.\ Eksp.\ Teor.\ Fiz.\  {\bf 98}, 462 (2004)]
  [arXiv:hep-ph/0111331].
  %%CITATION = ZETFA,98,462;%%

\bibitem{ESF2}
  A.~V.~Afanasev, I.~Akushevich and N.~P.~Merenkov,
  %``Radiative correction to the transferred polarization in elastic  electron
  %proton scattering,''
  Phys.\ Rev.\  D {\bf 65}, 013006 (2002)
  [arXiv:hep-ph/0009273].
  %%CITATION = PHRVA,D65,013006;%%

\bibitem{RADGEN}
I. Akushevich, H. Boettcher, D. Ryckbosch, In
Proc.
Workshop {\it ''Monte Carlo Generators for
HERA Physics'' (1998/99)}, Hamburg:DESY, (1999),
pp.~554-565.
\bibitem{POLRAD}
I. Akushevich, A. Ilyichev, N. Shumeiko, A. Soroko,
A. Tolkachev: Comput. Phys. Commun. 104 (1997) 201.
\bibitem{meradgen}
A. Afanasev, E. Chudakov,  A. Ilyichev,  and V. Zykunov,
Comput. Phys. Commun.  176 (2007) 218. 
\bibitem{mera}
A. Ilyichev and V. Zykunov,
Phys. Rev. D 72 (2005) 033018.

\bibitem{elradgen}
A.V. Afanasev, I. Akushevich, A. Ilyichev, B. Niczyporuk,
Czech.J.Phys. 53 (2003) B449.
\bibitem{hadvac}
H.~Burkhardt and B.~Pietrzyk,
  %``Update of the hadronic contribution to the QED vacuum polarization,''
  Phys.\ Lett.\  B 356, (1995) 398.
  %%CITATION = PHLTA,B356,398;%%

%\bibitem{praha2009}
%I.~Akushevich, A.~Ilyichev and N.~Shumeiko,
  %``ELRADGEN 2.0: Monte Carlo generator for simulation of radiative events in
  %polarized elastic electron-proton scattering,''
%arXiv:0912.2189 [hep-ph].
\bibitem{Ak} I.~Akushevich,
Eur.\ Phys.\ J.\  C 8, (1999) 457.
\bibitem{Doug}
D.~Hasell {\it et al.},
        %``The BLAST experiment,''
Nucl.\ Instrum.\ Meth.\  A 603, (2009) 247.
        %%CITATION = NUIMA,A603,247;%%
\bibitem{Tavi}
  O. Filoti, Ph.D. thesis, University of New Hampshire (2007).
\bibitem{Dre99}
  D. Drechsel \emph{et al.}, Nucl. Phys. A 645, (1999) 145.
\bibitem{AhR}
A. Akhiezer and M. P. Rekalo Sov. J. Part. Nucl. 4, (1974) 277.
\bibitem{FFunp}
I. A. Qattan \emph{et al.}: 
Phys. Rev. Lett. 94, (2005) 142301. 
\bibitem{FFpol1}
M. K. Jones \emph{et al.}
Phys. Rev. Lett. 84 (2000) 1398. 
\bibitem{FFpol2}
O. Gayou \emph{et al.}
Phys. Rev. Lett. 88, (2002) 092301. 
\bibitem{Gor}
T. Gorringe, H. W. Fearing, Rev. Mod. Phys. 76, (2004) 31. 
\bibitem{Beck}
D. H. Beck, R. D. McKeown,  Ann. Rev. Nucl. Part. Sci. 51, (2001) 189. 
\bibitem{HAPRAD}
I.~Akushevich, A.~Ilyichev and M.~Osipenko,
%``Radiative effects in the processes of hadron electroproduction,''
Phys. Lett. B 672, (2009) 35.

\end{thebibliography}
\end{document}